\documentclass[prx, preprint, groupedaddress]{revtex4-2}

\usepackage{graphicx}
\usepackage[normalem]{ulem}
\usepackage[final]{changes}
\usepackage{siunitx}
\usepackage{upgreek}

\usepackage{xcolor}
\usepackage{ulem}

\usepackage{lipsum}
\usepackage{amsmath}

\usepackage{ctable}

\usepackage{comment}

\usepackage{float}

\usepackage{scalerel,amssymb}

\begin{document}


\title{Measurement and Qualitative Explanation of Decay Lengths of Attractive and Repulsive Forces between Natural and Artificial Atoms}



\author{Marco Weiss}
\author{Fabian Stilp}
\author{Max Reinhart}
\author{Franz J. Giessibl}
\email{franz.giessibl@ur.de}
\affiliation{Institute of Experimental and Applied Physics, University of Regensburg, 93053 Regensburg, Germany}


\date{\today}

\begin{abstract}
Artificial atoms, such as quantum corrals, offer an excellent platform to study fundamental interactions between localized quantum states and nanoscale probes. We performed atomic force microscopy measurements inside square quantum corrals on Cu(111) using CO- and metal-terminated tips. Using chemically unreactive CO-terminated tips repulsive Pauli forces can be probed, while metallic tips are attracted to the localized quantum states due to chemical bonding.
We found distinct exponential decay constants of 46 pm for the repulsive and 66 pm for the attractive forces.
{\color{black}
Attractive and repulsive interactions between two natural atoms show significantly shorter decay lengths. While natural atoms feature states with a broad range of decay lengths, including very short ones from deeply bound states,  quantum corrals are lacking such deeply bound and highly localized states, resulting in longer decay lengths. 
These results offer a new route to understand and design atomic-scale interactions in low-dimensional quantum structures and devices.
}
\end{abstract}

\maketitle

\section{Introduction}
The ability to control and manipulate matter at the atomic scale is a fascinating aspect of modern nanoscience and nanotechnology.
A landmark achievement was the development of atomic manipulation \citep[][]{DM1990} using the scanning tunneling microscope (STM). Manipulating single atoms not only provides unprecedented visualization of surfaces \cite{Stroscio2004}, but also allows the construction of artificial quantum structures with dimensions all the way down to the nanometer scale. 
Atomic manipulation has proven to be remarkably versatile, allowing the exploration of fundamental quantum states in systems such as Dirac materials, including Lieb lattices \cite{Slot2017,Drost2017} and artificial graphene \cite{Gomes2012}, as well as in structures with interesting topological behavior \citep{Kempkes2019,Kempkes2023,Freeney2020_kekule}, just to name a fraction of what has been studied previously. It has also been applied to practical challenges, such as manipulating and sensing the spin state of artificial atomic structures (e.g., Refs. \cite{Hirjibehedin2006,Serrate2010,Esat2024}), logic operations \cite{Heinrich2002,Khajetoorians2011}, or pushing the limits of storing data at the atomic scale \cite{Moon2009,Kalff2016}.

By arranging adsorbates with atomic precision, researchers also gained the ability to tailor the confinement of electrons within these structures \cite{Crampin2000a,Schneider2023,Weiss2024}. Quantum corrals \cite{Crommie1993Corral}, formed by positioning atoms in a closed geometry on a surface, are a prime example. The closed geometry confines surface state electrons in the x-y-plane (surface-plane), leading to a set of resonant eigenstates inside the corral. 
By confining a fixed number of electrons in a small space such that a discretized energy spectrum is present, quantum corrals can be regarded as artificial atoms \cite{Kastner1993,Stilp2021}.
Since the original circular quantum corral, made from Fe adatoms on Cu(111), was introduced in 1993 \cite{Crommie1993Corral}, artificial atoms have been realized on a variety of more exotic surfaces, including semiconductors \cite{Sierda2023}, Rashba surface alloys \cite{Jolie2022a}, proximity superconductors \cite{Schneider2023}, and topological insulators \cite{Chen2019}. These structures have become essential model systems, allowing researchers to explore fundamental quantum phenomena in designed environments using STM.

Atomic force microscopy (AFM) has emerged as a complementary technique to STM, offering the ability to probe surface properties with atomic spatial resolution and femtonewton force sensitivity. 
While STM exclusively probes the electronic behavior of quantum structures such as artificial atoms, AFM allows for direct investigation of the forces that are determined by the interaction between the confined electronic states and the probe tip, an aspect STM does not provide. 
Understanding these interactions is vital for potential applications in quantum technology, such as quantum sensing or quantum computing.
Quantum sensors often rely on the interactions between the sensor and a probe (e.g., Refs. \citep{Degen2017,Heinrich2021}). Such sensors are very sensitive to even weak interactions, and the performance of such devices depends greatly on the understanding of how probes (e.g., the AFM tip) interact with the quantum system (e.g., the quantum corral). 
Quantum computing, on the other hand, requires precise control over quantum states (qubits) (e.g., Refs. \citep{DiVincenzo2000,Dowling2003,Hayashi2003,Williams,Yang2019}). Any unwanted or poorly understood interaction can lead to errors and decoherence. Therefore, a fundamental understanding of the forces that act on confined electronic states in quantum structures is important for the development of more robust and reliable qubits and quantum sensing devices.

AFM studies by Stilp \textit{et al.} \cite{Stilp2021} have demonstrated the ability to probe these tip-sample interactions within quantum corrals, revealing femtonewton-scale forces between the AFM tip and confined electronic states. Specifically, they observed a repulsive force when using a CO-terminated tip, attributed to Pauli repulsion, and an attractive force with a metal-terminated tip, suggesting the formation of a chemical bond. Although this study established the sensitivity of AFM to these interactions, no detailed discussion of the decay characteristics was given.

In this work, we present an AFM study of the interactions between the confined electronic state within square-shaped quantum corrals on a Cu(111) substrate and AFM tips with CO- and metal-terminations. Our research offers a detailed analysis of the attractive and repulsive forces, revealing distinct decay lengths of $46$ pm for repulsion and $66$ pm for attraction.
{\color{black} We further demonstrate that the measured force decays deviate form the conventional decay ratio found in the semi-empirical Morse-Potential, where there is a factor of 2 between the attractive and repulsive decay lengths. Based on our data, we estimate an upper limit of 1.6 for this factor. Additionally, a comparison of the decay lengths measured in a quantum corral with those observed between two natural atoms reveals a significant difference: Natural atoms typically exhibit decay lengths that are significantly shorter than the ones measured in a quantum corral. This can be attributed to the presence of deeper bound states in natural atoms, which contribute to interactions with a short decay length. Conversely, the quantum corral system lacks such deeply bound and highly localized states, which in turn leads to the longer decay lengths observed. This comparison highlights principles critical to understanding both the decay characteristics of interactions between natural atoms and the design of low-dimensional quantum structures with tailored chemical reactivity.}

\section{The quantum corral}
The quantum corral built for this work is square-shaped and consists of 26 individually positioned CO-Molecules. An STM topography image is shown in Fig. \ref{F1}(a). 
For details on the measurement conditions and experimental setup, see the Supplemental Material SM1 \cite{SupplMat}. 
\nocite{Giessibl1998,Albrecht1991,Emmrich2015,Welker2012,Welker2013,Wahl2008}
The side lengths of the quantum corral are $6.132$ nm (horizontal) and $6.195$ nm (vertical). Details about the exact adsorption sites of the CO molecules are provided in the Supplemental Material SM2 \cite{SupplMat}.
For the upper and lower corral walls, the inter-molecular distance is $1022$ pm, while the left and right walls have an inter-molecular spacing of $885$ pm.\\

\begin{figure}
    \centering
    \includegraphics[width=0.5\columnwidth]{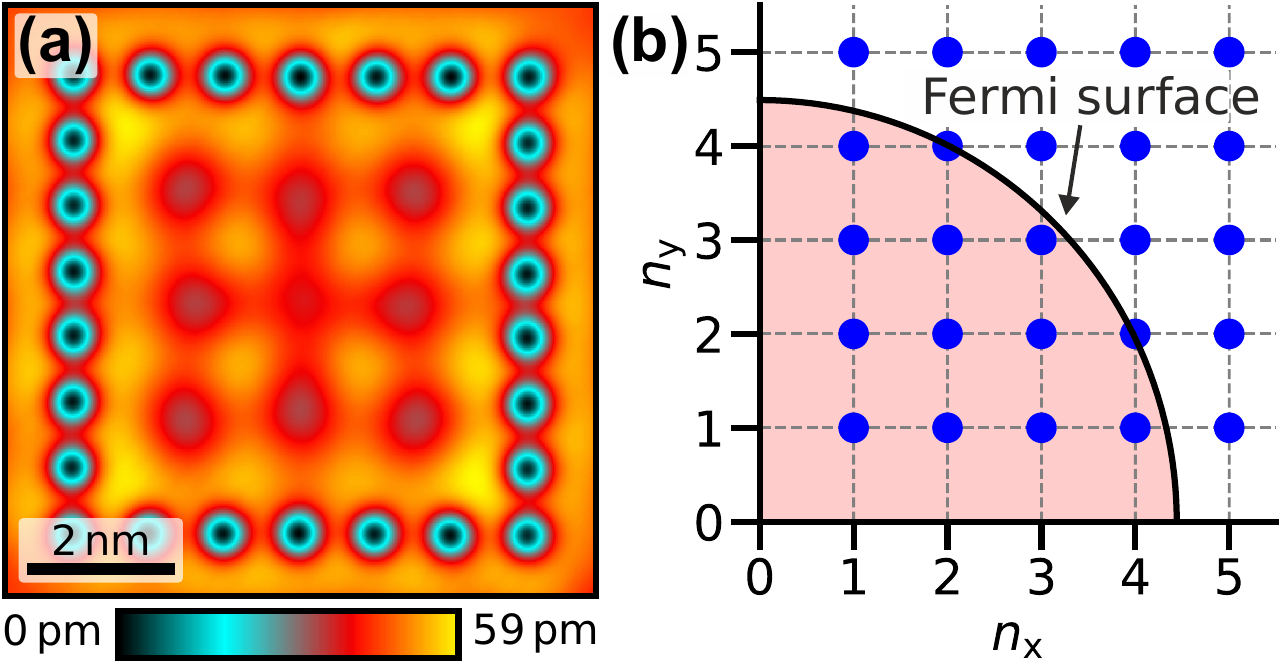}%
    \hfill
    \caption{\label{F1} (a): STM topography image of the square corral, consisting of 26 CO-molecules, measured with a metal tip, a sample bias of $10$ mV, and a tunneling current setpoint of $100$ pA. The side lengths are $6.132$ nm (horizontal) and $6.195$ nm (vertical). CO-molecules appear as dark dips. In the inner part of the corral, the charge density near the Fermi-level can be seen as the checkerboard-like structure. (b): Model-spectrum of the square shaped corral in (a) in $n$-space, characterized by the main quanum numbers $n_\mathrm{x}$ and $n_\mathrm{y}$. All states (blue markers) which lie in the red shaded region are below the Fermi level and are therefore occupied with electrons.} 
\end{figure}

A quantum corral on Cu(111) confines the quasi-free 2D electron gas present on the surface (also called the Shockley surface state \cite{Shockley1939}), resulting in a set of resonant eigenstates \cite{Crommie1993Corral}. 
To understand the behavior of the confined Shockley surface state within the quantum corral, a model description of a particle in a 2D box can be used. It was shown that an infinitely high potential well in the $x$- and $y$-plane (surface-plane) captures the main properties of a quantum corral, including the spatial shape and energetic position of the resonant eigenstates \cite{Crommie1993Corral,Stilp2021,Weiss2024}.
In the $x$- and $y$- directions there are sinusoidal-shaped standing waves characterized by the quantum numbers $n_\mathrm{x}$ and $n_\mathrm{y}$. These quantum numbers determine the energy and shape of the confined electron wave functions $\Psi_\mathrm{n_\mathrm{x}, n_\mathrm{y}}(x, y)$. 
For explanatory purposes, a modeled spectrum of the quantum corral in $n$-space is given in Fig. \ref{F1}(b). Each marker represents a single corral state, which is characterized by its main quantum numbers $n_\mathrm{x}$ and $n_\mathrm{y}$. All corral states below the Fermi level (within the red-shaded region in Fig. 1(b)) are occupied with electrons. A more detailed discussion about the model-description of the quantum corral is provided in the Supplemental Material SM3 \cite{SupplMat}.

Stilp \textit{et al.} were the first to show that the measured forces between the AFM-tip and the Shockley surface state confined in a quantum corral scale with the total surface charge density $\sigma_\mathrm{tot.}$ \cite{Stilp2021}.
The total surface charge density is determined by which corral states lie below the Fermi energy $E_\mathrm{F}$ and are therefore occupied by electrons: $\sigma_\mathrm{tot.}(x,y) \propto \sum_{E_{n_\mathrm{x}, n_\mathrm{y}} 	\leq E_\mathrm{F}} |\Psi_\mathrm{n_\mathrm{x}, n_\mathrm{y}}(x, y)|^2$. Metal-tips showed a stronger chemical attraction to regions with an increased total surface charge density, whereas CO-terminated tips are repelled from regions with an increased surface charge density. This repulsive force was attributed to Pauli repulsion \cite{Stilp2021}.

The decay of the corral states towards the vacuum ($z>0$) is not affected by the lateral confining structure (the walls) and is described by $\Psi_z (z) \propto \mathrm{exp}(-z/ \lambda_\mathrm{S}^\Psi )$ with $\lambda_\mathrm{S}^\Psi$ being the decay constant of the sample wave functions. 
Calculations by Stilp \textit{et al.}  \cite{Stilp2021} showed that corral states decay with $\lambda_\mathrm{S}^\Psi = 84$ pm towards the vacuum.
Consequently, the corresponding electron density ($|\Psi|^2$) decays with a length of $42$ pm.

\section{Experiments and model description}
\subsection{Constant height AFM images and the total surface charge density}\label{IIIA}

\begin{figure}
    \includegraphics[width=0.55\columnwidth]{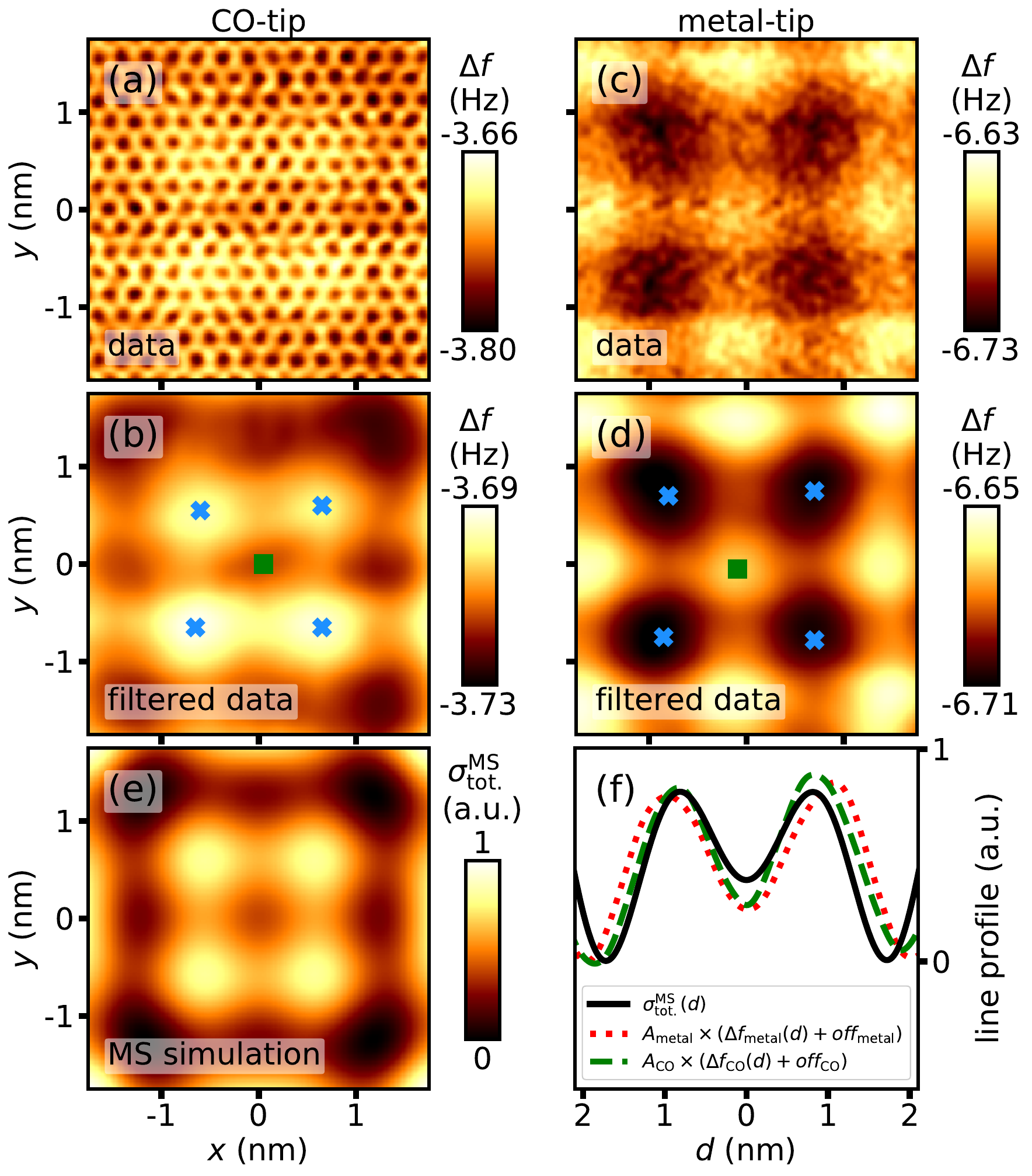}%
	\caption{\label{F2} (a): Frequency shift ($\Delta f$) image measured with a CO-tip in constant height showing the Cu surface atoms as small dark dips. The confined surface state is visible as the superimposed four bright spots around the center surrounded by eight light dark dips. 
    (b): Gaussian low-pass filtered version of (a) to remove the Cu grid \cite{Stilp2021,SupplMat}. 
    (c): Frequency shift image measured with a metal-tip in constant height. The confined surface state is visible as four dark dips around the center surrounded by eight bright spots. 
    (d): Slightly Gaussian low-pass filtered version of (c). 
    (e): Multiple scattering simulation of the total surface charge density  $\sigma_\mathrm{tot}^\mathrm{MS}$. The appearance of (b) and (d) only differs by a minus sign, showing that local accumulations of the surface charge density lead to repulsion when measured with a CO-tip (bright regions in (b)), and attraction (dark regions in (d)), when measured with a metal-tip.
    { \color{black}
    (f): Average of the two diagonal line profiles of the CO-tip AFM image from (b), shown as the green dashed line. In this plot $d$ is defined as the distance from the center of the corral. For qualitative comparison, an offset of $off_\mathrm{CO} = 3.73$ Hz was added and the result multiplied by an amplitude factor of $A_\mathrm{CO} = 31\,\frac{\mathrm{a.u.}}{\mathrm{Hz}}$: $A_\mathrm{CO}\times(\Delta f_\mathrm{CO} + off_\mathrm{CO})$.
    The red dotted line shows the corresponding profile from the metal-tip AFM image in (d), scaled by using $off_\mathrm{metal} = 6.65$ Hz and $A_\mathrm{metal} = -14.7\,\frac{\mathrm{a.u.}}{\mathrm{Hz}}$.
    The black line represents the simulated profile of (e).
    This comparison illustrates that both AFM images reflect the total surface charge density $\sigma_\mathrm{tot}^\mathrm{MS}$. } 
    }
\end{figure}

In Fig. \ref{F2}(a) a constant height AFM image of the inner part of the quantum corral is provided, measured with a CO-terminated tip and a sample bias of $V_\mathrm{B} = 0$ V. This image was taken at a tip-sample distance of $340$ pm, defined by the core-core distance of the frontmost oxygen atom and the copper surface layer \cite{Schneiderbauer2014}.
The measurement signal $\Delta f$ is proportional to the force gradient between tip and sample \cite{Giessibl2001}.
At such a small tip-sample separation, the Cu surface atoms are visible as small attractive (dark) features \cite{Schuler2013,Emmrich2015}. Overlaid with this copper grid, wider features can be recognized. This additional signal stems from the confined surface state and consists of four bright spots around the center of the image surrounded by eight light-dark dips. Applying a Gaussian low-pass filter (see SM4 \cite{SupplMat} for more details) to Fig. \ref{F2}(a) suppresses the spatially high-frequency atomic surface lattice signal and shows the confined surface state more clearly (see Fig. \ref{F2}(b)).  

When AFM measurements are performed with a metal-tip (see Figs. \ref{F2}(c) and \ref{F2}(d)) the opposite behavior is found. Metal-tips interact attractively with corral states, resulting in an increased attraction over regions with a high $\sigma_\mathrm{tot.}$ (dark areas). 

To simulate the total surface charge density, multiple scattering (MS) simulations are employed.
The MS formalism developed by Heller \textit{et al.} \cite{Heller1994} is based on an s-wave scattering approach and is an established method to simulate the electronic properties of a quantum corral. It has been shown that the best agreement between experiment and MS simulation can be achieved when the $\delta$-peak shaped scattering potentials are provided with a strong absorbing channel \cite{Heller1994}, which is called ``black-dot'' limit. Our simulations showed that the confined surface state electrons exhibit a slightly higher effective mass $m_\mathrm{e, conf.}^*$ than the free surface state electrons $m_\mathrm{e, free}^*$. 
The best agreement between simulation and measurement was found for $m_\mathrm{e, conf.}^* = 0.45\times m_\mathrm{e}$, while the effective mass of the free Shockley surface state is $m_\mathrm{e, free}^* = (0.405\pm0.025) \times m_\mathrm{e}$ (e.g., Refs. \cite{Crommie1993,Gartland1975,Courths2001,Forster2003,Hyldgaard2003,Butti2005}). The slight increase in effective mass of confined electrons is a known phenomenon that has also been observed, for example, in semiconductor quantum dots \cite{Bekhouche2018}. Furthermore, Freeney \textit{et al.} \cite{Freeney2020} also simulated square-shaped quantum corrals on Cu(111) with a muffin tin model and found effective masses of $0.48\times m_\mathrm{e}$ and $0.46\times m_\mathrm{e}$, which is in good agreement with the effective mass we found. The effective mass was determined to ensure a correct simulation, and consistency between simulation and experiment, but it is not central to our analysis and will not be discussed further. Additional details on the MS simulations are given in the Supplemental Material SM5 \cite{SupplMat}.

The simulated total surface charge density $\sigma_\mathrm{tot.}^\mathrm{MS}$ is shown in Fig. \ref{F2}(e). In this image, bright regions correspond to a high total surface charge density, while dark regions correspond to a low total surface charge density.
Comparison of Figs. \ref{F2}(b) and \ref{F2}(e) again makes it clear that AFM measurements, conducted with a CO-terminated tip, are sensitive to the total surface charge density, where an increased total surface charge density (bright regions in Fig. \ref{F2}(e)) results in an increased repulsive force (bright areas in Fig. \ref{F2}(b)). 
The opposite behavior can be found when comparing the metal-tip AFM image (Figs. \ref{F2}(c) and (d)) with the simulation of the total surface charge density (Fig. \ref{F2}(e)). An increased total surface charge density (bright regions in Fig. \ref{F2}(e)) results in a more attractive force (dark in Fig. \ref{F2}(d)). Regions with a minimum of total surface charge density (dark in Fig. \ref{F2}(e)) will result in a minimal attractive force, which appears bright in AFM images (see bright regions in Fig. \ref{F2}(d)).
In Fig. \ref{F2}(f) the average line profiles along the two diagonals of the respective images are shown. To show qualitative agreement, an offset was added to all measured line profiles and then multiplied by an amplitude factor. Details are provided in the caption of Fig. \ref{F2}(d). Please note that the line profile of the metal-tip AFM image (red dotted line) was multiplied with a negative amplitude factor, effectively inverting the profile line.

{\color{black} 
It is worth noting that the maxima of the profile lines in Fig. \ref{F2}(d) are further apart for the metal-tip measurement than for the CO-tip measurement. For the metal-tip the maxima are separated by 2 nm while for the CO-tip they are separated by 1.7 nm.
Approaching an AFM-tip close to the surface can shift the energy of the corral states \cite{Stilp2021}. It was shown that CO-tips slightly raise the energy of corral states, leading to a small depopulation of corral states near the Fermi energy. In contrast, metal-tips lower the state energies, slightly increasing the electron occupation. These tip-induced shifts affect the total surface charge density and can explain the subtle differences between the CO- and metal-tip profile lines.
This confirms the interpretation of the tip-induced energy shifts shown by Stilp \textit{et al.} \cite{Stilp2021} in circular quantum corrals.
A more detailed discussion is provided in the Supplemental Material SM6 \cite{SupplMat}.
Multiple scattering simulations do not account for such a tip-induced energy shift. 
The calculated $\sigma_\mathrm{tot}^\mathrm{MS}$ might also deviate moderately from the measurements as -440 meV for the band edge of the Shockley surface state was used, a value that varies slightly in literature.
}

\subsection{Decay lengths of the attractive and repulsive forces}
\begin{figure}
    \includegraphics[width=0.7\columnwidth]{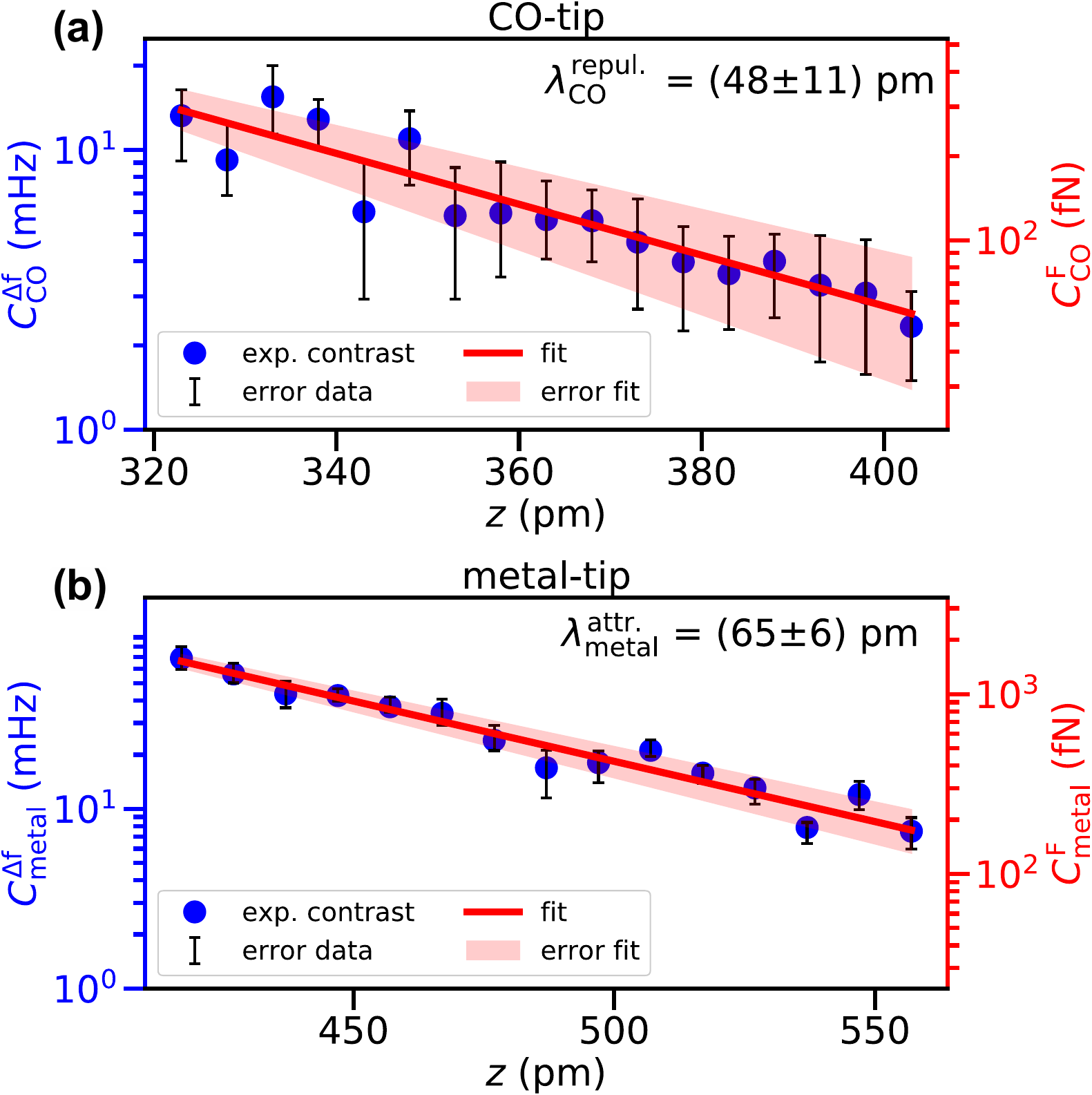}%
	\caption{\label{F3} (a): Contrast in $\Delta f$, measured with a CO-tip, as a function of vertical distance $z$ showing an exponential behavior with a decay length of $\lambda_\mathrm{CO}^\mathrm{repul.} = (48 \pm 12)$\;pm. The contrast is defined as the difference between the average of the four innermost maxima and the central minimum. (b): Contrast in $\Delta f$, measured with a metal-tip, with a decay length of $\lambda_\mathrm{metal}^\mathrm{attr.} = (64 \pm 7)$\;pm. The contrast is defined as the difference between the central maximum and the average of the four innermost minima. 
    The definition of the error bars of the data points is described in the Supplemental Material SM7 \cite{SupplMat}.   
    In both cases the $\Delta f$-contrast was fitted with an exponential decay function. The fits are displayed as full red lines and can be associated to a force contrast (right axis). The red shaded region gives the error of the fit which is two times the standard deviation.}
\end{figure}

Recording constant height AFM images at different tip-sample separations additionally provides insight into the distance dependence of the interaction between the tip and the corral states. Figure \ref{F3} shows the contrast evolution of the frequency shift $C^\mathrm{\Delta f}$ as a function of the tip-sample distance $z$. Each $C^\mathrm{\Delta f}$-value in Fig. \ref{F3} was calculated from one constant height AFM image. The contrast is defined by the absolute difference between the four innermost extrema, marked by blue crosses, and the center, marked by the green square (see Figs. \ref{F2}(b) and (d)). More details about the calculation of the contrast is provided in the Supplemental Material SM7 \cite{SupplMat}.

For both tip configurations (CO- and metal-tip), the contrast decays exponentially with $z$, resulting in decay lengths of $\lambda_\mathrm{CO}^\mathrm{repul.} = (48 \pm 11)$ pm and $\lambda_\mathrm{metal}^\mathrm{attr.} = (65 \pm 6)$ pm. The force contrast $C^\mathrm{F}$ follows from equation (8) in \cite{Giessibl2019}: $C_\mathrm{CO}^\mathrm{F}(z) = 291\;\mathrm{fN} \times \mathrm{exp}(- (z-323\;\mathrm{pm}) / \lambda_\mathrm{CO}^\mathrm{repul.})$ for the CO-tip measurements and $C_\mathrm{metal}^\mathrm{F}(z) = 1522\;\mathrm{fN} \times \mathrm{exp}(-(z-417\;\mathrm{pm}) / \lambda_\mathrm{metal}^\mathrm{attr.} )$ for the metal-tip measurements. This means that for the smallest tip-sample separations ($323$ pm for CO- and $417$ pm for metal-tip) force contrasts of $291\;\mathrm{fN}$ and $1522\;\mathrm{fN}$ were measured. The respective force contrasts are shown as full red lines in Fig. \ref{F3}. 
For reproducibility, additional measurements were made using different CO- and metal-tips in a separate corral \cite{footnote3}. The corresponding analysis is provided in the Supplemental Material SM8 \cite{SupplMat} and is summarized in Table \ref{Table}. 
It is evident that the obtained decay lengths are consistent with the measurement accuracy. This demonstrates that the decay lengths are independent of the mesoscopic tip geometry and the environment of the quantum corral.

\begin{figure}
    \includegraphics[width=\columnwidth]{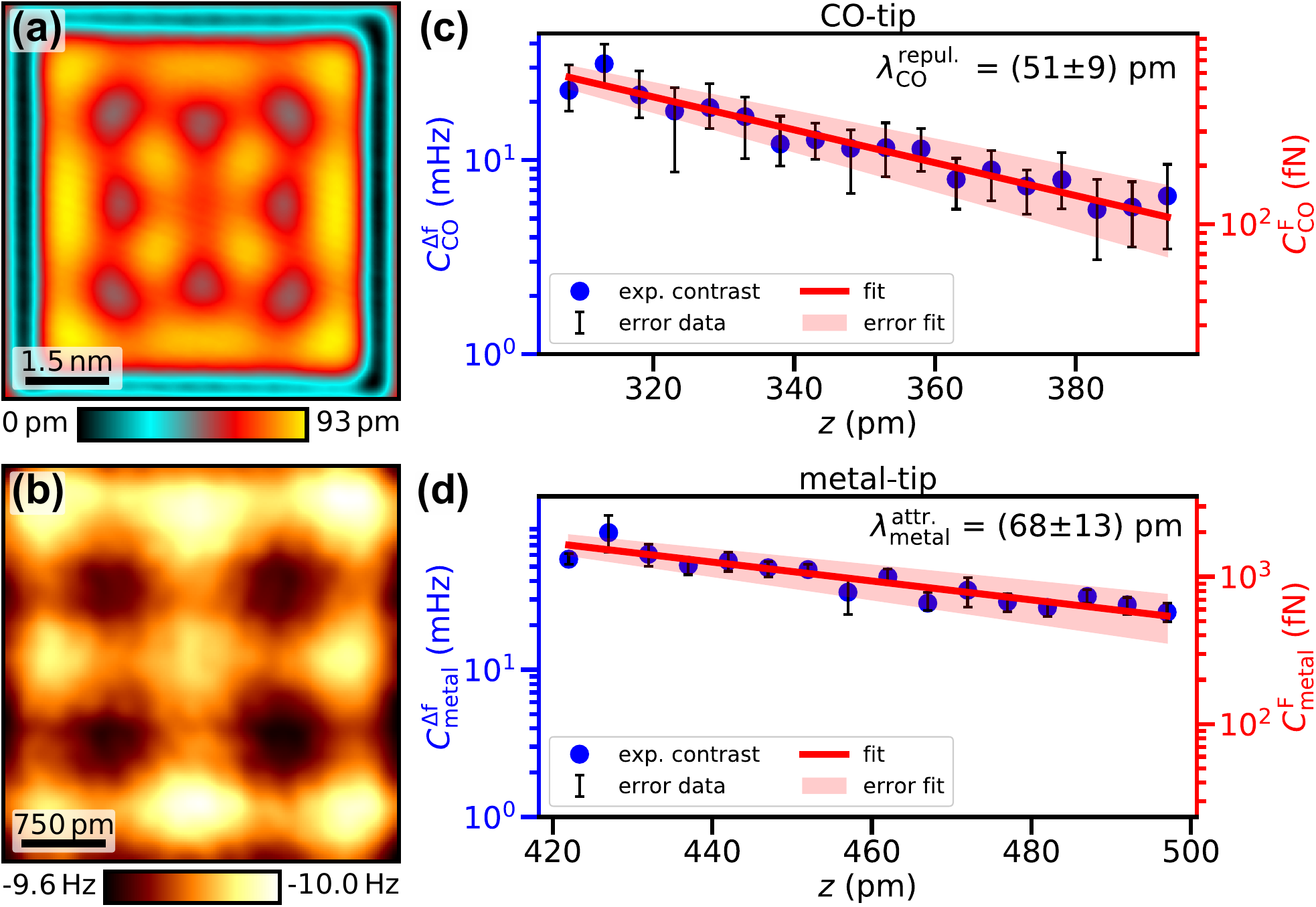}%
	\caption{\label{F4}(a): STM topography image of the denser square corral measured with a metal tip, a sample bias of $-10$ mV, and a tunneling current setpoint of $100$ pA. This corral also has side lengths of $6.132$ nm and $6.195$ nm but consits of $52$ CO-molecules. (b): Frequency shift image measured with a metal-tip in constant height. (c) and  (d): Contrast in $\Delta f$ and $F$, measured with a CO- and a metal-tip, as a function of vertical distance $z$ showing an exponential behavior with decay lengths of $\lambda_\mathrm{CO}^\mathrm{repul.} = (51 \pm 9)$\;pm and  $\lambda_\mathrm{metal}^\mathrm{attr.} = (68 \pm 13)$\;pm.}
\end{figure}

Recently, it was demonstrated that the eigenstates of quantum corrals are influenced by the wall density. Higher wall densities cause a slight shift in the energies of the corral states toward higher values and also extend the lifetimes of these states \cite{Weiss2024}. However, the decay of the corral states into the vacuum should not be affected by the wall density. To show this, AFM measurements were also performed in a corral with twice the wall density compared to the one presented in Fig. \ref{F1}(a). The decay lengths of the denser quantum corral are listed in Table \ref{Table} and the corresponding analysis is shown in Fig. \ref{F4}.

As shown in Table \ref{Table} and Fig. \ref{F4} the measured decay lengths remain unchanged within the measurement accuracy, regardless of the mesoscopic tip configuration or the wall density of the corral. This demonstrates a clear relationship between the decay lengths of the interaction of the confined Shockley surface state with a metal- and a CO-tip (attractive and repulsive):
\begin{equation}\label{eq:relation}
    \lambda^\mathrm{attr.}_\mathrm{metal} = (1.4 \pm 0.2) \times \lambda^\mathrm{repul.}_\mathrm{CO}.
\end{equation}

\begin{table}
    \centering
    \begin{tabular}{c || c  c }
         & \quad $\lambda_\mathrm{CO}^\mathrm{repul.}$ (pm) \quad & \quad $\lambda_\mathrm{metal}^\mathrm{attr.}$ (pm) \quad \\ \hline \hline
        corral 1 \qquad & $48 \pm 11$ & $65 \pm \;\;6$ \\ 
        (Fig. \ref{F1} to \ref{F3}) & & \\ \hline
        corral 2 \qquad & $40 \pm 13$ & $67 \pm \;\;8$ \\
        (SM8 \cite{SupplMat}) & $39 \pm 14$  & \\ \hline
        denser corral \qquad& $51 \pm \;\;9$ & $68 \pm 13$ \\
        (Fig. \ref{F4}) & & \\  \specialrule{.2em}{.05em}{.05em} 
        average \qquad & $46 \pm \;\;6$ & $66 \pm \;\;5$ \\
    \end{tabular}
    \caption{\label{Table} AFM force contrast decay lengths measured with a CO- and a metal-tip. Corral 1 was measured using a monoatomic metal-tip, followed by measurements with a CO-terminated tip (same metal-tip but equipped with a CO at the apex). Corral 2, built after several preparation cycles on the same copper crystal, was measured using a different mesoscopic tip configuration \cite{footnote3}. To verify the reproducibility of the decay lengths, measurements with the CO-tip were repeated twice. Corral 3, which featured denser walls, was measured using the same CO- and metal-tip as for corral 1.}
\end{table}

Equation (\ref{eq:relation}) shows that in the case of the corral, the decay lengths of Pauli repulsion and chemical bond differ by a factor of $1.4$. This result shows a clear difference between the decay lengths of the attractive and repulsive interactions with a confined Shockley surface state.

{\color{black}
As discussed earlier in Section \ref{IIIA} and SM6 \cite{SupplMat}, the close proximity of the AFM tip can slightly shift the center energy of the corral states. This alters their occupation and, consequently, the total surface charge density (see the difference between attractive and repulsive profile lines in Fig. \ref{F2}(f)). In general, such changes could influence the distance dependence of the $\Delta f$-contrast. Corral states in denser structures have longer lifetimes and therefore narrower energy distributions \cite{Weiss2024}, making their total surface charge densities more sensitive to tip-induced shifts. However, since both dense and less dense corrals yield the same decay length (see Table \ref{Table}), we conclude that the influence of this effect on the measured decay length is negligible.
}

\subsection{Model description of the attractive and repulsive forces}

The extracted decay length of the CO-tip measurement is greater than the decay length of the surface charge density ($\lambda_\mathrm{S}^\Psi/2 = \lambda_\mathrm{S}^\rho = 42$ pm). This can be understood with a commonly used model description of Pauli repulsion, which is given by the overlap-integral of the total charge densities of the tip $\rho_\mathrm{CO-tip}$ and sample $\rho_\mathrm{sample}$: 
\begin{equation}\label{eq:Pauli}
    F_\mathrm{Pauli, z} \propto \frac{\partial}{\partial z} \int [\rho_\mathrm{CO-tip}(\vec{r}) \rho_\mathrm{sample}(\vec{r})]^\alpha \mathrm{d} \vec{r},
\end{equation}
with $\alpha \approx 1$ (e.g., Refs. \cite{Kim1981, Moll2012,Ellner2019,Liebig2020}). From this relationship, it becomes evident that the measured force contrast conducted with a CO-tip in quantum corrals depends not only on the decay length of the total charge density of the sample but also on that of the tip. As an estimate, the charge density at the apex of the CO-tip can be modeled as a s-wave like distribution (i.e., Slater-type functions for large distances from the atomic nucleus). With an effective nuclear charge $Z_\mathrm{eff} \approx 4.5$ of the valence electrons of oxygen \cite{Slater1930} the charge density of the front most part of the CO-tip approximately decays with $12$ pm. With this the model description of Pauli repulsion gives a decay length of the force of $\lambda_\mathrm{Pauli, z}^\mathrm{model} = 43$ pm  with $\alpha = 1$ which is slightly larger than the charge density decay of the Shockley surface state (see Supplemental Material for more details SM9.1 \cite{SupplMat}). The modeled decay length also lies within the measured error window of $46\pm 6$ pm.

{\color{black}
Despite the good agreement, we found two possible reasons why the modeled decay length of Pauli repulsion is slightly shorter than the measured length. None of those is the s-wave approximation of the flat corral states, as this is justified in SM9.5 \cite{SupplMat}.

One possibility lies in the fact that the Shockley surface state decay length was calculated using the known work function of Cu(111). The close proximity of the AFM tip can locally alter the potential landscape of the surface, effectively modifying the work function \cite{Lang1988,Ternes2011}. Such a perturbation could lead to a longer decay length than expected from ideal, unperturbed surface parameters. An effective work function of $\Phi^\mathrm{CO-tip}_\mathrm{eff.} \approx 4.5$ eV is needed to closely match the measured decay lengths.

Another contributing factor may be the scaling exponent $\alpha$, which appears in the density-overlap-based model of Pauli repulsion \cite{Kim1981,Ellner2019}. Although set to unity in this work for clarity and consistency, it was shown that $\alpha$ can vary depending on the nature of the interacting systems. Values slightly above 1 have been reported for molecular systems \cite{Ellner2019}, while interactions between single natural atoms often yield values slightly below 1 \cite{Kim1981}. A value of $\alpha = 0.92$ results in a modeled decay length that closely matches the measured length. This suggests that modest deviations from the idealized scaling behavior may be intrinsic to the corral system and should be taken into account in future refinements of the model.
}

The distance dependence of chemical bonding is often approximated by the overlap of wave functions (e.g., Refs. \cite{Baym, Ternes2011, Jelinek2012, Nordholm2020}): $F_\mathrm{chem, z} \propto \frac{\partial}{\partial z} \int \Psi_\mathrm{tip}(\vec{r}) \Psi_\mathrm{sample}(\vec{r}) \mathrm{d} \vec{r}$. In the relevant energy regime around the Fermi level, metallic tips are predominantly characterized by s-orbitals \cite{Gustafsson2017} with an estimated decay length of $\approx 84$ pm \cite{PHD_Andi,footnote4}. 
This simplified model predicts a chemical force decay length of $125$ pm (see Supplemental Material for details SM9.2 \cite{SupplMat}).

With this model description of the distance dependence of a chemical bond, the key difference between Pauli repulsion and chemical attraction is that the former depends on the absolute square of the wave functions ($\rho \propto |\Psi|^2$), while the latter depends on the wave function itself ($\Psi$). Because the charge densities of tip and sample decay as $|\Psi|^2$, the decay length of Pauli repulsion is inherently shorter than that of the chemical attraction. This trend is also reflected in our measurements, where Pauli repulsion results in a shorter decay length (CO-tip, $46$ pm) than chemical bonding (metal-tip, $66$ pm).

It is important to note that the chosen description of the chemical force between the metal-tip and corral states is a simplified model. Although it qualitatively explains the observed difference in decay lengths, it results in a decay length twice as long as the measured one. A more sophisticated approach, such as a linear combination of atomic orbitals (LCAO) model, would inherently incorporate orbital overlap effects in a more detailed manner. However, the development of a complete LCAO-based description goes beyond the scope and purpose of this work.

{\color{black}
The commonly used semi-empirical Morse-potential \cite{Morse1929} provides a useful and intuitive model to describe the potential energy of diatomic molecules. It consists of two summed exponential functions, one describing the repulsive interaction (prominent for small separations) and one describing the attractive interaction (prominent for large separations). The Morse-potential has been widely applied in various fields of science, ranging from the description of catalytic processes on surfaces to the simulation of dynamics in biological systems or gases \cite{Mirzanejad2025}. Given this broad applicability, it also serves as a meaningful reference for AFM-based force analysis.

In the original Morse formulation, the attractive decay length is twice as long as the repulsive one, meaning $\mu = \lambda_\mathrm{attr.}^\mathrm{Morse} / \lambda_\mathrm{repul.}^\mathrm{Morse} = 2$. However, this ratio cannot be applied to the measurements presented here. Our data, recorded with a metal-tip, clearly lie in the attractive regime which corresponds to the right side of the potential minimum. Attempting to access the repulsive regime was not feasible, as the strong forces between tip and sample caused tip deformations or even dropping tip-material inside the corral. Therefore, the metal-tip measurements were limited to the attractive tail of the interaction, and fitting a full Morse-potential was not possible.

Instead, we approximated the interaction by fitting a single exponential function to the attractive side (see Figs. \ref{F3} and \ref{F4}), which yields a decay length of $66$ pm. However, fitting one side of a Morse-like potential with a single exponential function introduces a systematic overestimation. The steep repulsive term slightly distorts the shape of the attractive tail, which is not captured in such a simplified fit. A mathematical analysis (see SM10 \cite{SupplMat}) shows that the fitted decay length $\lambda_\mathrm{metal}^\mathrm{attr.}$ is always greater than or equal to the true attractive decay length: $\lambda_\mathrm{metal}^\mathrm{attr.} \geq \lambda_\mathrm{real}^\mathrm{attr.} $. This implies that $\lambda_\mathrm{real}^\mathrm{attr.} \leq 66$ pm.
If the original Morse ratio $\mu = 2$ is applied, this would suggest a decay length of the Pauli repulsion of $\lambda_\mathrm{real}^\mathrm{repul.} \leq 33$ pm. As shown for the CO-tip measurements, Pauli repulsion in the corral is well described by the overlap integral of electron densities (see Eq. (\ref{eq:Pauli})), and the decay length of the surface state charge density $\lambda_\mathrm{S}^\mathrm{\rho} = 42$ pm can be regarded as a lower bound for the repulsive decay length for the corral system (see SM9.4 \cite{SupplMat}). This directly excludes the Morse factor of $2$. Taking the measured attractive decay length as an upper limit, the maximum possible ratio is: $\mu_\mathrm{max} = 66\,\mathrm{pm}/42\,\mathrm{pm} = 1.6$. While this does not contradict the general concept of the Morse-potential, it shows that the specific decay ratio assumed in its original form is not applicable to the experimental conditions in this study.
}

{\color{black}
A rough estimate of the repulsive decay length of the metal-tip can be obtained by applying the same model used for the CO-tip (electron density overlap in Eq. (\ref{eq:Pauli})). While the repulsive regime could not be accessed directly in the experiment, this method yields a hypothetical decay length of approximately 
$52$ pm (see SM9.3 for further details \cite{SupplMat}). Taking this value as the relevant repulsive decay length, the upper bound of the decay length ratio reduces to $\mu_\mathrm{max} = 66\,\mathrm{pm}/52\,\mathrm{pm} = 1.3$. This defines an even stricter upper bound for the decay length ratio, further supporting that the original Morse ratio of $2$ is incompatible for the quantum corral system.
}

{\color{black}
\subsection{Comparison of an artificial atom to a natural atom}

To better understand the measured decay lengths of tip–corral interactions, it is instructive to compare them with those observed in interactions between natural atoms. Similar to the artificial atom case, natural systems also show that chemical bonding generally decays slower than Pauli repulsion. However, the absolute decay lengths observed in natural atom–atom interactions are shorter than those found in the corral system.

For chemical bonding, Ternes \textit{et al.} \cite{Ternes2011} reported decay lengths around $40$ pm for interactions between a metal-tip and individual Cu and Pt adatoms on Cu(111). Rubio-Bollinger \textit{et al.} \cite{Bollinger2004} found similar values for metallic adhesion between gold atoms in atomic junctions, with decay lengths of approximately $45$ pm. In contrast, the attractive interaction measured in this work between a metal-terminated AFM tip and the corral states shows a much longer decay length of $66$ pm.

A comparable trend holds for Pauli repulsion. Measurements with CO-terminated tips on CaF\textsubscript{2} yielded repulsive decay lengths of $27$ pm \cite{Liebig2020}, and atomic scattering experiments report values around $28$ pm across a range of various atomic species \cite{Inouye1972,Inouye1972a,Inouye1973,Kita1975,Kita1975a,Inouye1979}. In contrast, the CO-tip measurement over the quantum corral yields a repulsive decay length of $46$ pm.

A likely explanation for this consistent difference lies in the nature and energy range of the electronic states involved in the interaction. Natural atoms have an energetically wide-spread distribution of electronic states, extending from the Fermi level down to deep-lying core levels. These states span several eV below $E_\mathrm{F}$ and include orbitals that are highly localized in space. For instance, in Cu(111), the sp-bands begin about $10$ eV below the Fermi level and are energetically broad. The 3d-band starts around $5$ eV below $E_\mathrm{F}$ and is nearly fully occupied. This is sketched on the lower left side of Fig. \ref{F5}(a).

\begin{figure}
    \includegraphics[width=\columnwidth]{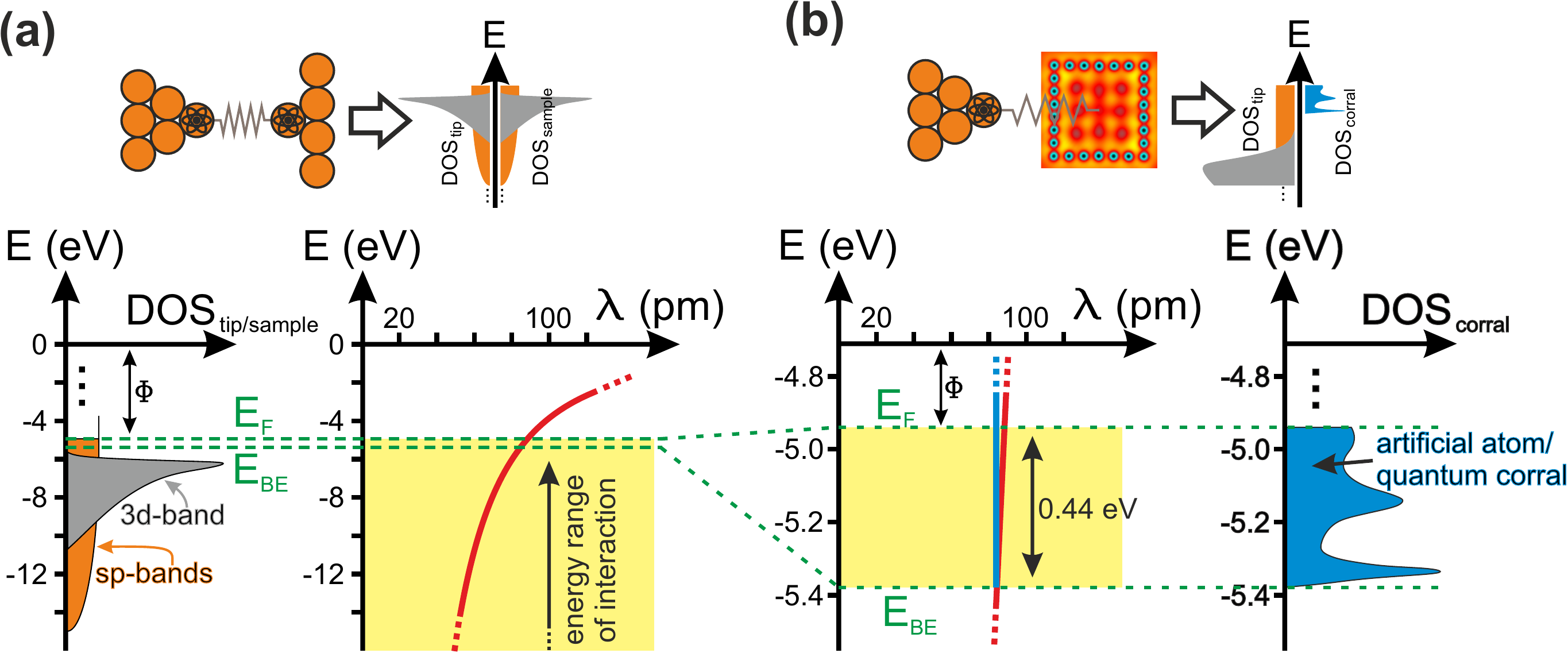}%
	\caption{\label{F5}
    {\color{black} Qualitative sketch of the electronic states involved in tip–sample interactions.
    (a) Bottom left: Schematic density of states (DOS) for a natural atom at the AFM tip apex and a natural adatom on a surface. For illustrative purposes, both DOS profiles (for tip and sample) are assumed to be identical. Tip and sample exhibit broad sp- and d-bands. Lower lying bands are not shown. $E_\mathrm{F}$ is the Fermi energy, which is $\Phi = 4.94$ eV below zero energy and $E_\mathrm{BE}$ gives the band edge of the Shockley surface state.    
    Bottom Right: Approximated decay length $\lambda$ of electronic states in a natural atom configuration as a function of energy, following a $\lambda \propto 1/\sqrt{-E}$ dependence (see main text).
    (b) Electronic interaction between a natural atom and an artificial atom (quantum corral). The relevant energy range for interaction is defined by $E_\mathrm{BE}$ and $E_\mathrm{F}$. Within this narrow energy window the tip-side decay lengths vary only slightly (between $84$ pm and $87$ pm), while the decay length of the corral states remains constant at $84$ pm (see graph at the bottom left side). 
    Bottom right: Local density of states in the center of the corral (blue), extracted from a $\mathrm{d}I/\mathrm{d}V$ spectrum (see SM5 \cite{SupplMat}).
    Top right: Comparison of the DOS profiles of tip and quantum corral. The DOS profile of the tip is a zoomed in version of the natural atom profiles shown in (a).
    Averaged over the energy range of interaction the natural-natural atom interaction has a shorter decay length as the natural-artificial atom interaction.
    }
    }
\end{figure}

The energetic dependence of the decay length of states, can be roughly approximated with an electronic state in a one-dimensional box. With the energy $E$ relative to the vacuum level ($=0$) the exponential decay length is given by $\lambda = \hbar/\sqrt{-2 m_\mathrm{e} E}$. For example, the 1s core level of copper, with a binding energy of roughly $9000$ eV \cite{Nordling1957}, has an approximated decay length of $2$ pm. With an atomic spacing of $255$ pm in bulk copper, such deeply bound orbitals barely overlap, and thus, can be considered to not form bands. In contrast, decay lengths at the onset of the Shockley surface state ($440$ meV below $E_\mathrm{F}$) are $84$ pm. The energy dependence of the approximated decay lengths is shown on the lower right side of Fig. \ref{F5}(a).

The total chemical interaction between two natural atoms is governed by contributions from many orbitals across a wide energy range. Long ranging orbitals near the Fermi level decay slowly and begin to overlap at relatively large distances. As atoms come closer, deeper, more localized orbitals start to contribute, resulting in a faster increase in force. The full interaction can thus be seen as a combination of many orbital overlaps, each with its own distance dependence. This combined interaction decays faster than the Fermi-level component alone, which explains why experimental decay lengths for natural atoms are often significantly shorter than those of the highest-lying states. This is also the reason for the excellent spatial resolution of AFM \cite{Welker2012}.

The situation in the artificial atom (quantum corral) is different. The occupied electronic states of the corral lie within a narrow energy range of $440$ meV below the Fermi level, as illustrated on the lower right side of Fig. \ref{F5}(b). For the quantum corral there are no deeply bound and spatially highly localized states contributing to the interaction. 
Because of the 2D nature of the corral states, the decay lengths of all corral states are 84 pm.
Within the narrow energy range of the occupied corral states, the approximated energy dependent decay lengths of electronic states of natural atoms vary only slightly (from $84$ pm to $87$ pm).
This is depicted on the lower left side of Fig. \ref{F5}(b). The interaction is therefore governed solely by slowly decaying states, resulting in longer decay lengths for both repulsive and attractive interactions compared to natural atom–atom systems.

That bonding in natural systems often involves deeper-lying states is also supported by several examples from literature. The 3d-band of noble metals are known to play a critical role in the adsorption of CO, O, and CO\textsubscript{2} on copper clusters \cite{Li2012}, as well as on transition metals like Cu, Ag, and Au \cite{Saini2022}. This band is centered several eV below the Fermi level. In other cases, even fully filled semicore orbitals were shown to participate in bonding. For instance, the existence of HgF\textsubscript{4} is explained by bonding contributions from Hg’s filled d-orbitals \cite{Wang2007}, and CsF\textsubscript{5} has been predicted to form with bonding contributions from fully filled p-orbitals \cite{Rogachev2015}. Another example is the uranyl ion UO$_{2}^{2+}$, where the main bonding orbitals are 5f and 6d, but at short bond lengths, even the filled 6p semicore orbitals hybridize and contribute to the interaction \cite{Walch1976,Tatsumi1980,Pyykko2000}. It was further shown that also in solids, deeper-lying atomic orbitals participate in bonding \cite{Orlov2022}. These examples underline that bonding in natural systems can involve orbitals located well below the Fermi level, further supporting the conclusion that the effective decay length in such systems can be shortened by the participation of fast-decaying, deep-lying states.
}

\section{Summary and Outlook}
\subsection{Summary}
We performed constant-height AFM measurements with CO- and metal-terminated tips within a square-shaped quantum corral on Cu(111). Multiple scattering simulations supported the observation of previous studies \cite{Stilp2021} that the forces measured with AFM in quantum corrals scale with the total electron density. CO-terminated tips interact repulsively with the confined surface state, while metal-terminated tips interact attractively. We further found that the forces measured with CO- and metal-terminated tips decay exponentially with different decay lengths. The construction of a quantum corral with the same geometry but a denser wall confirmed that the decay lengths are independent of the corral wall and the mesoscopic tip shape. The repulsive interaction (CO-tip) exhibited a decay length of $\lambda_\mathrm{CO}^\mathrm{repul.} = (46 \pm 6)$ pm, while the attractive interaction (metal-tip) showed a decay length of $\lambda_\mathrm{metal}^\mathrm{attr.} = (66 \pm 5)$ pm. These decay lengths differ by a factor of $ 1.4 \pm 0.2$.  

Modeling the repulsive interaction between a CO-tip and the quantum corral states using an established model for Pauli repulsion (electron density overlap integral) yielded a modeled decay length of $\lambda_\mathrm{model}^\mathrm{repul.} = 43$ pm. Modeling the chemical attraction between a metal-tip and the corral states using a wave function overlap integral showed the same trend as the measurements ($\lambda_\mathrm{metal}^\mathrm{attr.} > \lambda_\mathrm{CO}^\mathrm{repul.}$), but resulted in a too long decay length.

{\color{black} 
Due to experimental constraints, a full Morse-potential could not be fitted to the metal-tip measurements. Instead a single exponential was used to fit the attractive side, introducing a systematic overestimation of the attractive decay length ($\lambda_\mathrm{metal}^\mathrm{attr.} \geq \lambda_\mathrm{real}^\mathrm{attr.}$).
Analysis showed that the original Morse-potential's decay ratio of attractive to repulsive forces ($\mu = \lambda_\mathrm{attr.}^\mathrm{Morse} / \lambda_\mathrm{repul.}^\mathrm{Morse} = 2$) is not applicable to the corral system. Instead, the maximum possible ratio was found to be $\mu_\mathrm{max} = 1.6$. This demonstrates that the specific decay ratio assumed by Morse is not applicable to the experimental conditions of the quantum corral.
}

A comparison between natural and artificial atoms further reveals that the decay lengths of both attractive and repulsive interactions are significantly longer in the quantum corral system. This is attributed to the narrow energy window of occupied corral states near the Fermi level, which lack contributions from short-ranged, deep-lying orbitals. In contrast, natural atoms involve a broad spectrum of electronic states, including rapidly decaying orbitals far below $E_\mathrm{F}$, leading to a shorter effective decay length.

\subsection{Outlook}

In their AFM study of circular quantum corrals, Stilp \textit{et al.} were the first to reveal that confined surface states can establish chemical interactions with a metallic AFM tip, and they discovered attractive decay lengths of $50$ pm and $56$ pm \cite{Stilp2021}. 
The discrepancy in decay lengths between our results (square shaped quantum corral, $66$ pm) and their circular corral points to the possibility that the corral geometry influences the decay length and strength of chemical interactions. Further investigation of this relationship may show how the corral wave functions can be shaped to tailor the chemical reactivity of confined Shockley surface states.

Artificial nanostructures further offer an intriguing framework for investigating new electrical and chemical capabilities, going beyond single quantum corrals. Sierda \textit{et al.} \cite{Sierda2023} developed a method for simulating the electronic structure of molecules using small, coupled quantum corrals on InSb. Building on this, it would be interesting to see if these simulated molecules have a similar chemical reactivity compared to their natural counterparts.

The emerging field of designer electronic structures has already produced fascinating systems with unique properties, such as Lieb lattices and artificial graphene \cite{Slot2017, Drost2017,Gomes2012} with Dirac cones, or SSH chains and kagome lattices with interesting topological behavior \cite{Kempkes2019,Kempkes2023}. There, the natural question arises: What bonding characteristics do these exotic electronic states exhibit and do a CO- and a metal-tip interact with these exotic states in a manner similar to that with a single quantum corral? 

Moreover, quantum corrals were already built on non-traditional materials such as topological insulators \cite{Chen2019}, proximity superconductors \cite{Schneider2023}, semiconductors \cite{Sierda2023}, graphene \cite{Lee2016}, or Rashba surface alloys \cite{Jolie2022a}. Within these frameworks, investigating the chemical bonding interactions of corral states to a probe-tip, may present a promising direction for future exploration.

All of these potentially interesting artificial nanostructures can also be measured with different tip terminations. Established tip terminations include, among others, Xe \cite{Eigler1991,Neu1995}, copper oxide \cite{Monig2016,Monig2018}, or even magnetic molecules such as nickelocene \cite{Verlhac2019}. Performing experiments similar to those presented here with different tip terminations could drastically increase the understanding of the nature of chemical bonds in a controlled environment with tunable chemical reactivity. Also spin-polarized tips \cite{Wiesendanger1990,Bode2003} could be used to investigate bonding properties of corral states on surfaces where magnetism plays a crucial role (e.g., corrals on topological insulators or superconductors).

\begin{acknowledgements}
The authors thank A. Weindl and C. Setescak for carefully proofreading the manuscript and are grateful to A. Donarini for many fruitful discussions.
\end{acknowledgements}

%


\pagebreak

\renewcommand\thetable{\arabic{figure}}
\renewcommand{\tablename}{Table}

\renewcommand\thefigure{S\arabic{figure}}
\renewcommand{\figurename}{Figure}

\renewcommand\thesection{SM\arabic{section}}
\renewcommand\thesubsection{SM\arabic{section}.\arabic{subsection}}

\setcounter{equation}{0}
\setcounter{figure}{0}
\setcounter{table}{0}
\setcounter{section}{0}
\setcounter{subsection}{0}

\begin{center}
	\LARGE \textbf{Supplemental Material}\\
\end{center}

\section{Experimental setup}
The experiments were carried out using a home-built UHV microscope at a temperature of $5.6$ K. The microscope is equipped with a qPlus sensor \cite{Giessibl1998} with a stiffness of $k = 1800$ N/m, a resonance frequency of $20,382$ Hz, and a Q-factor of $163,000$. 
STM measurements were performed with the bias voltage $V_\mathrm{B}$ applied to the sample.
At STM topography images the tip-sample distance was modulated to achieve a constant current.
For AFM measurements, the frequency modulation technique \cite{Albrecht1991} was used, with all measurements performed at an oscillation amplitude of $A = 50$ pm. 
The tungsten tip was prepared by poking it into the Cu(111) surface until it ended in a single front atom, which was verified using the COFI (CO front-atom identification) method \cite{Emmrich2015,Welker2012,Welker2013}. The presented AFM measurements were first performed with this metal-tip, which was then terminated with a CO molecule, ensuring the same tip background for both tip configurations. 
The Cu(111) sample was prepared by multiple Ar-sputtering and annealing cycles.

\newpage

\section{Corral construction plans} 

\begin{figure}[H]
	\centering
	\includegraphics[width=0.49\textwidth]{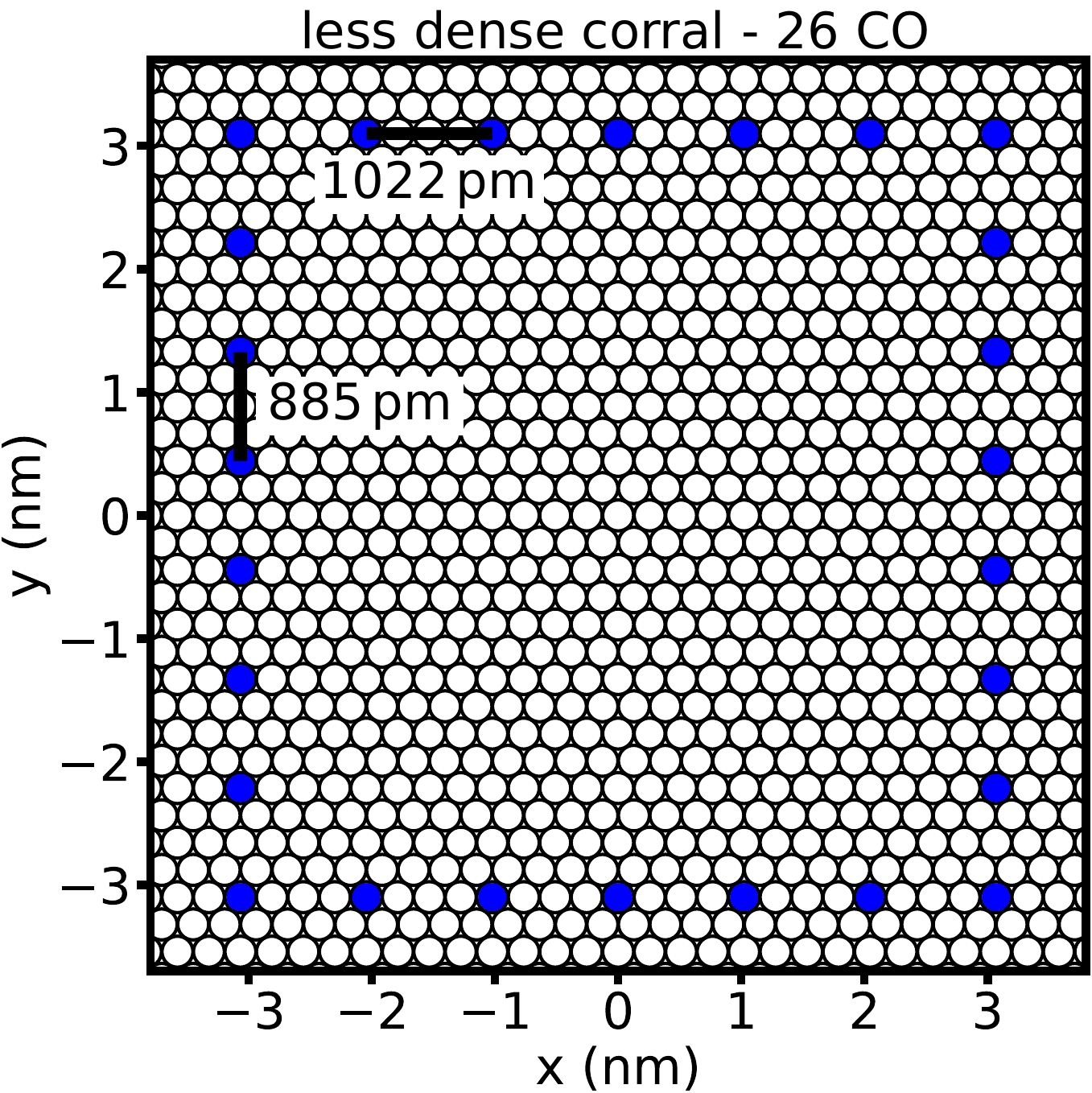}%
	\hfill
	\includegraphics[width=0.49\textwidth]{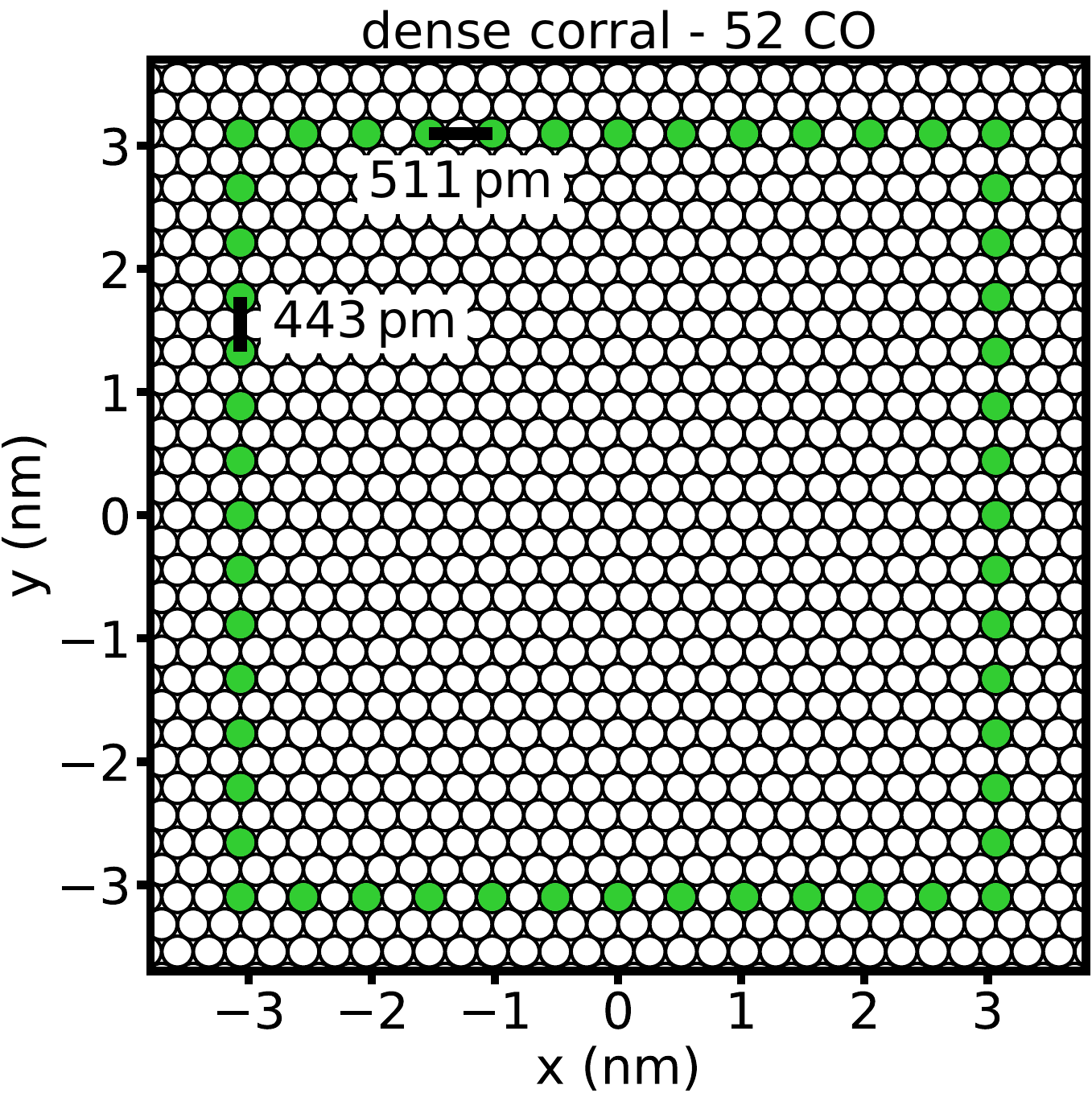}%
	\caption{\label{fig:construction_plan} Adsorption sites for the two corrals. Unfilled black circles are the Cu(111) surface atoms. The filled colored circles give the adsorption position of the CO molecules. The less dense quantum corral consists of 26 CO molecules and the denser corral consists of 52. Both have the same size. The side lengths are $6.132$ nm (horizontal) and $6.195$ nm (vertical). } 
\end{figure}

In Figure \ref{fig:construction_plan} the construction plans of the corrals are displayed. CO molecules on Cu(111) adsorb on the top site. The atomic spacing of the copper surface atoms is $255.5$ pm. For constructing the dense quantum corral, we took the building plan of the less dense corral and inserted additional CO molecules.

\newpage

\section{Model description of the quantum corral}\label{sec:model_corral}

Like mentioned in the main text it was shown that an infinitely high potential well in the x- and y-plane (e.g. hard wall model) captures the main properties of a quantum corral, including the spatial shape and energetic position of the resonant eigenstates. Solving the Schrödinger equation for a rectangular quantum well in the surface plane gives the wave functions:

\begin{equation}\label{HW_wave function}
\Psi_{n_\mathrm{x}, n_\mathrm{y}} (x, y) = N \times \mathrm{sin}\biggl( \dfrac{\pi n_\mathrm{x} x}{L_\mathrm{x}} \biggr) \times \mathrm{sin}\biggl( \dfrac{\pi n_\mathrm{y} y}{L_\mathrm{y}} \biggr).
\end{equation}

The corral wave functions, $\Psi_{n_\mathrm{x}, n_\mathrm{y}}(x,y)$, are characterized by the main quantum numbers $n_\mathrm{x}$ and $n_\mathrm{y}$, while the side lengths of the corral are  given by $L_\mathrm{x}$ and $L_\mathrm{y}$. The corral states satisfy the normalization condition $\int\int |\Psi_{n_\mathrm{x}, n_\mathrm{y}}|^2 \mathrm{d}A = 1$ with the normalization constant $N$:
\begin{equation}
N = \sqrt{\dfrac{4 \pi n_\mathrm{x}}{L_\mathrm{x}[\mathrm{sin}(2 \pi n_\mathrm{x}) - 2\pi n_\mathrm{x}] } \times \dfrac{4 \pi n_\mathrm{y}}{L_\mathrm{y}[\mathrm{sin}(2 \pi n_\mathrm{y}) - 2\pi n_\mathrm{y}] } }.
\end{equation}
The eigenenergies of the system are given by:
\begin{equation}\label{HW_engergy}
E_{n_\mathrm{x}, n_\mathrm{y}} = \dfrac{\hbar^2 \pi}{2 m_\mathrm{e}^*} \biggl(\dfrac{n_\mathrm{x}^2}{L_\mathrm{x}^2} + \dfrac{n_\mathrm{y}^2}{L_\mathrm{y}^2} \biggr),
\end{equation}

where $\hbar$ is the reduced Plank's constant and $m_\mathrm{e}^*$ is the effective mass of the surface state electrons. Equation (\ref{HW_engergy}) makes it clear that for every combination of $n_\mathrm{x}$ and $n_\mathrm{y}$ the belonging corral state $\Psi_{n_\mathrm{x}, n_\mathrm{y}}(x,y)$ is located at a specific energy. The energy in this model description starts at the onset of the Shockley surface state band, which is located $440$ meV below the Fermi level. Corral states with $E_{n_\mathrm{x}, n_\mathrm{y}} \leq 440$ meV are therefore occupied by electrons. The Fermi surface in two dimensional $n$-space representation is then given by:

\begin{equation}
E_F = 440\;\mathrm{meV} = \dfrac{\hbar^2 \pi}{2 m_\mathrm{e}^*} \biggl(\dfrac{n_x^2}{L_x^2} + \dfrac{n_y^2}{L_y^2} \biggr)
\end{equation}
\begin{equation}
\rightarrow n_y= \sqrt{ \dfrac{ 2 m_\mathrm{e}^* L_y^2 \times 440\;\mathrm{meV}}{\hbar^2 \pi} - \dfrac{L_y^2}{L_x^2}  n_x^2 } .
\end{equation}

\newpage

\section{Suppression of the copper grid}
By measuring with a CO-terminated tip inside the quantum corral, the AFM is sensitive to both, the confined Shockley surface state and the Cu surface atoms. These appear as attractive (dark) spots in CH AFM scans (see Fig. 2(a) in the main text).

To enhance the visibility of the AFM signal originating from the confined Shockley surface state, we applied a Gaussian low-pass filter to suppress the spatially high-frequency Cu grid signal. Following a similar approach like used in \cite{Stilp2021}, we first transformed the AFM image into frequency space using a fast Fourier transformation (FFT). By multiplying the FFT image with a two-dimensional Gaussian kernel (in the $k_\mathrm{x}$ and $k_\mathrm{y}$ directions), we reduced the spatially high-frequency components. Finally, the filtered FFT image was transformed back into real space. This process was performed twice in succession: FFT $\rightarrow$ Gaussian filtering $\rightarrow$ inverse FFT, followed by a second round of the same procedure. The sequential procedure enables better control over the minimum required filter parameters.

The Gaussian filter parameters correspond to the pixel width (full width at half maximum) in the $k_\mathrm{x}$ and $k_\mathrm{y}$ directions. All recorded AFM images contain between $(192\times 192)$ pixel and $(208\times 208)$ pixel. For measurements with a CO-tip, filtering was necessary to suppress the Cu grid signal, and images were processed twice using a Gaussian kernel with a pixel width of 15 pixels. In contrast, AFM images taken with a metal tip did not show the Cu grid and therefore did not require suppression of high-frequency grid features. Instead, they were filtered purely to smooth the images for easier processing, using the same two-step procedure with a Gaussian kernel of 20 pixels.

\newpage

\section{Multiple scattering simulation and effective mass}
Previous studies on semiconductor quantum dots \cite{Bekhouche2018} have demonstrated that the effective electron mass, $m_\mathrm{e}^*$, varies with the size of the quantum structure. Smaller quantum dots were found to exhibit a larger effective electron mass, suggesting that $m_\mathrm{e}^*$ depends on the physical dimensions or strength of the confining structure.

\begin{figure}[H]
	\centering
	\includegraphics[width=0.49\textwidth]{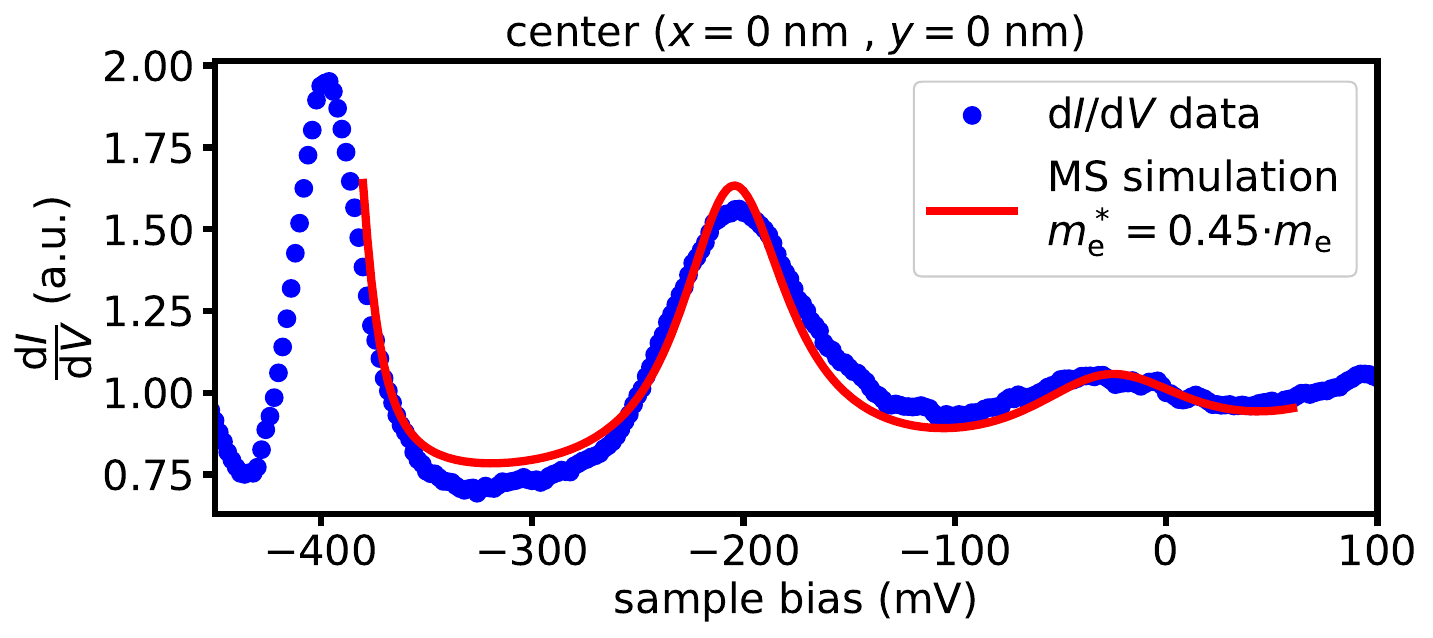}%
	\hfill
	\includegraphics[width=0.49\textwidth]{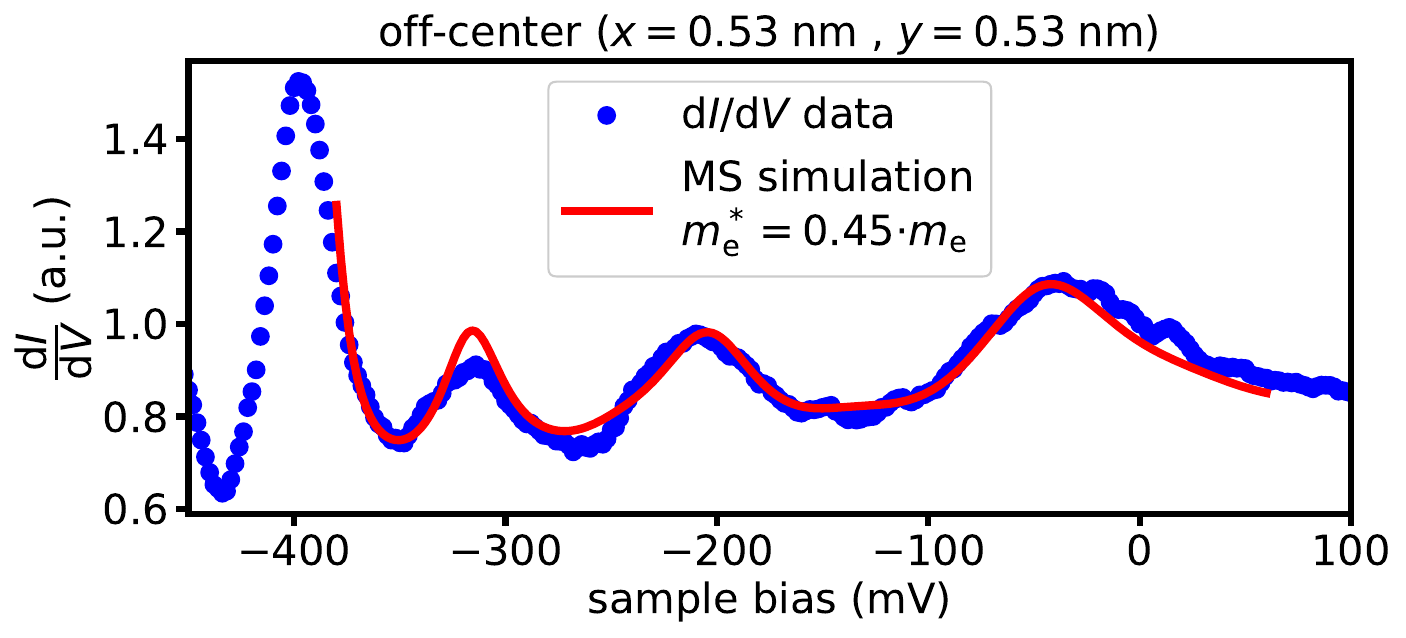}%
	\caption{\label{effektiveMasse} Comparison of d$I/$d$V$ spectra (blue markers) and multiple scattering (MS) simulations (full red line). The best agreement between data and simulation were found at an effektive electron mass of $m_\mathrm{e}^* = 0.45 \times m_\mathrm{e}$. } 
\end{figure}

To determine $m_\mathrm{e}^*$ in our structure, we performed multiple scattering simulations in the black-dot limit and compared the results with differential conductance measurements ($\mathrm{d}I/\mathrm{d}V$) in the corral. These measurements were performed at two locations within the corral: one at the center (x = 0 nm, y = 0 nm) and one off-center at coordinates (x = 0.53 nm, y = 0.53 nm). The measurements are presented in Figure \ref{effektiveMasse}. To remove the remaining influence of the tip's density of states (DOS) from the measurements, the sample's local density of states (LDOS) was obtained using the full deconvolution method by Wahl \textit{et al.} \cite{Wahl2008}.

Our analysis revealed that the best agreement between data and multiple scattering simulations was achieved with an effective mass of $m_\mathrm{e}^* = 0.45 \times m_\mathrm{e}$. Comparing this value with the known values for the free Shockley surface state on Cu(111), $m_\mathrm{e}^* = [0.38,0.43] \times m_\mathrm{e}$, indicates that the effective mass in the confined structure is larger than in the free case. Here $m_\mathrm{e}$ is the mass of a free electron.

The value of $m_\mathrm{e}^*$ found in this study is close to previously reported values of $0.48$ and $0.46$ \cite{Freeney2020}. In that work, also square corrals on Cu(111) were investigated, and the reported values were calculated using a muffin-tin model.

With the used multiple scattering formalism it is possible to simulate the variation of the local density of states ($LDOS(x, y, \epsilon)$) depending on the energy $\epsilon$ of the surface state electrons.
Integrating the simulated $LDOS$ from the onset of the surface band up to the Fermi-level then gives a measure of the total surface charge density $\sigma_\mathrm{tot.}$.

\newpage

{\color{black}
	\section{AFM-tip induced change in the total surface charge density}

	To evaluate how a tip-induced shift in the central energy of corral states affects the total surface charge density, we use the corral wave functions and energies described in section \ref{sec:model_corral}.
	
	Because corral states have finite lifetimes, they are broadened in energy and do not appear as sharp delta peaks. Instead, each state exhibits a Gaussian energy distribution \cite{Weiss2024}. 
	A simple model based on a single parameter, the average path length $x$, has been shown to describe the width $\eta$ of these Gaussian distributions well. For example, a circular corral with a diameter of $7.13$ nm exhibits an average path length of $x \approx
	5$ nm \cite{Weiss2024}. Given that the corral in this study has a comparable size and wall density, we adopt the same value for our simulations.
	
	Each energy distribution $G$ is described by:
	\begin{equation}
	G(E, E_{n_x, n_y}) = \dfrac{1}{\sqrt{2\pi\times\eta(E_{n_x, n_y})^2}} \mathrm{exp}\biggl(- 0.5 \dfrac{(E-E_{n_x, n_y})^2}{\eta(E_{n_x, n_y})^2} \biggr),
	\end{equation}
	
	where $E$ is the energy (starting from the Shockley surface state band edge), $E_{n_x, n_y}$ is the center energy of the corral state, and $\eta(E_{n_x, n_y})$ is the width parameter of the distribution. The width depends on the energy of the respective state and is given by \cite{Weiss2024}:
	\begin{equation}
	2\sqrt{2 \mathrm{ln}(2)} \times \eta(E_{n_x, n_y}) = \dfrac{\hbar \sqrt{\dfrac{2E_{n_x, n_y}}{m_\mathrm{e}^*}}}{x}.
	\end{equation}
	Integrating $G$ up to the Fermi level ($440$ meV above the band edge) gives the occupation $p_{n_x, n_y}$ of each state. Corral states far below the Fermi level are fully occupied ($p = 1$), while those near the Fermi energy are only partially filled ($0 < p < 1$). The total surface charge density is then constructed as a weighted sum:
	\begin{equation}
	\sigma_\mathrm{tot}^\mathrm{HW}(x,y) \propto \sum_{n_x, n_y} p_{n_x, n_y}\times|\Psi_{n_x, n_y}(x,y)|^2.
	\end{equation}
	
	To simulate the effect of the tip, we shift the energy of all corral states downward by $15$ meV. This leads to an increased occupation for states near the Fermi level, as a larger portion of their energy distribution now lies below the Fermi energy. The occupations of states far below the Fermi energy, which are already fully occupied, remain unaffected. A summary of the unshifted and shifted energies, width parameters, and occupations is given in Table \ref{tab:summary_filling}. For example, the state with  $n_x = 1$ and $n_x = 1$ shows no change in occupation, while a state closer to the Fermi energy (such as $n_x = 2$, $n_x = 4$) become more populated after the tip-induced shift.
	
	In Fig. \ref{fig:diff_occu}(a), we show line profiles of the total surface charge densities (along the diagonal of the corral) for the unshifted and shifted cases. A zoom-in of the central region is shown in Fig. \ref{fig:diff_occu}(b). 
	The maxima of the red curve (shifted states) are further apart than the maxima of the blue curve (unshifted states). This is consistent with the  increased contribution from partially filled states.
	The maxima of the red curve are separated by 2.0 nm while for the blue curve they are separated by 1.9 nm. 
	This effect closely mirrors the difference observed in the AFM line profiles measured with CO- and metal-tips in Fig. 2(f) of the main text.
	
	\begin{table}[h]
		\centering
		\begin{tabular}{cc|ccc|ccc}
			\hline
			$n_x$ & $n_y$ & $E_{n_x, n_y}$ (meV) & $\eta$ (meV) & $p$ & $E_{n_x, n_y, s}$ (meV) & $\eta_s$ (meV) & $p_s$ \\
			\hline
			1 & 1 & 44.0  & 10.4 & 1.00 & 29.0  & 8.4  & 1.00 \\
			1 & 2 & 109.3 & 16.3 & 1.00 & 94.3  & 15.2 & 1.00 \\
			2 & 1 & 110.7 & 16.4 & 1.00 & 95.7  & 15.3 & 1.00 \\
			2 & 2 & 176.0 & 20.7 & 1.00 & 161.0 & 19.8 & 1.00 \\
			1 & 3 & 218.2 & 23.1 & 1.00 & 203.2 & 22.3 & 1.00 \\
			3 & 1 & 221.8 & 23.3 & 1.00 & 206.8 & 22.5 & 1.00 \\
			2 & 3 & 284.9 & 26.4 & 1.00 & 269.9 & 25.7 & 1.00 \\
			3 & 2 & 287.1 & 26.5 & 1.00 & 272.1 & 25.8 & 1.00 \\
			1 & 4 & 370.6 & 30.1 & 0.99 & 355.6 & 29.5 & 1.00 \\
			4 & 1 & 377.3 & 30.4 & 0.98 & 362.3 & 29.8 & 1.00 \\
			3 & 3 & 396.0 & 31.1 & 0.92 & 381.0 & 30.5 & 0.97 \\
			2 & 4 & 437.3 & 32.7 & 0.53 & 422.3 & 32.1 & 0.71 \\
			4 & 2 & 442.7 & 32.9 & 0.47 & 427.7 & 32.3 & 0.65 \\
			3 & 4 & 548.4 & 36.6 & 0.00 & 533.4 & 36.1 & 0.00 \\
			4 & 3 & 551.5 & 36.7 & 0.00 & 536.5 & 36.2 & 0.00 \\
			\hline
		\end{tabular}
		\caption{\color{black}Center energies $E_{n_x, n_y}$, width parameters $\eta$, and occupations $p$ of corral states before and after a simulated tip-induced energy shift of $-15$ meV. Subscript $s$ denotes the shifted values.}
		\label{tab:summary_filling}
	\end{table}
	
	\begin{figure}[H]
		\centering
		\includegraphics[width=0.49\textwidth]{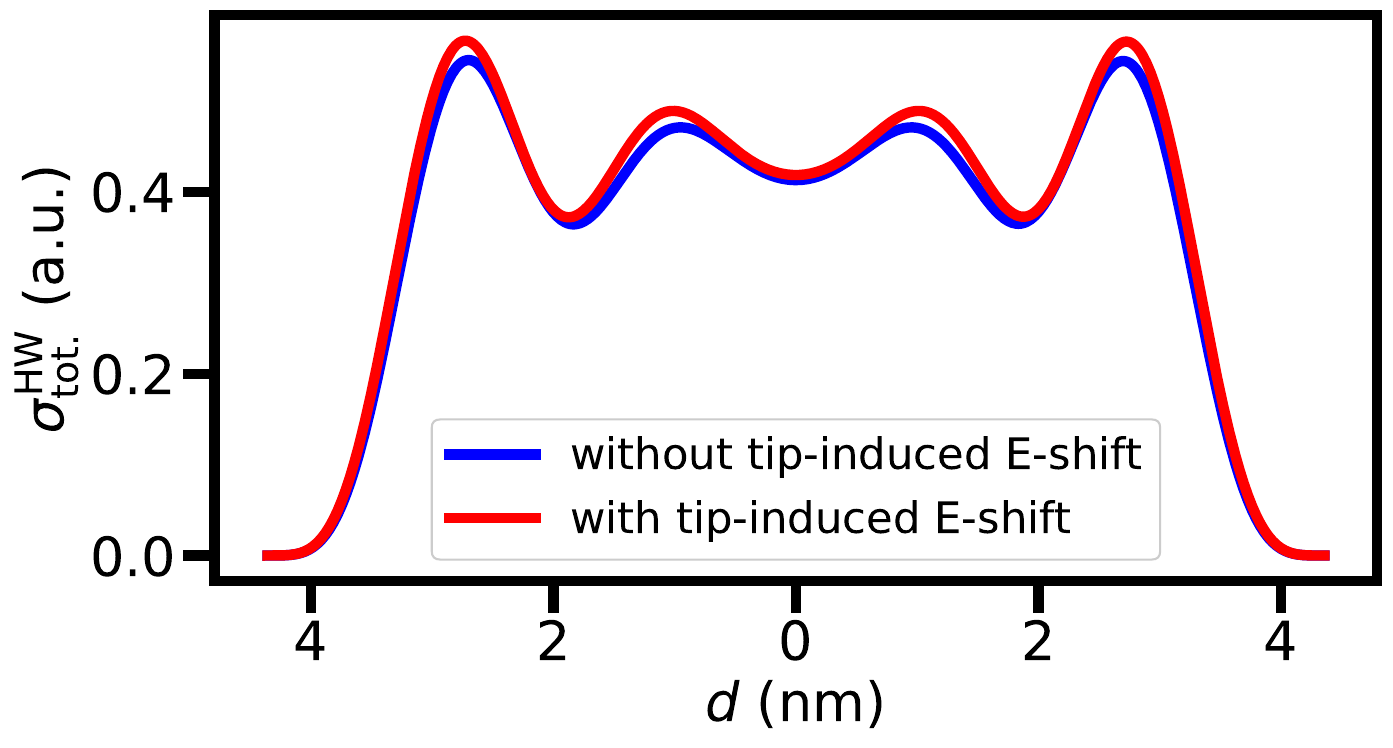}%
		\hfill
		\includegraphics[width=0.49\textwidth]{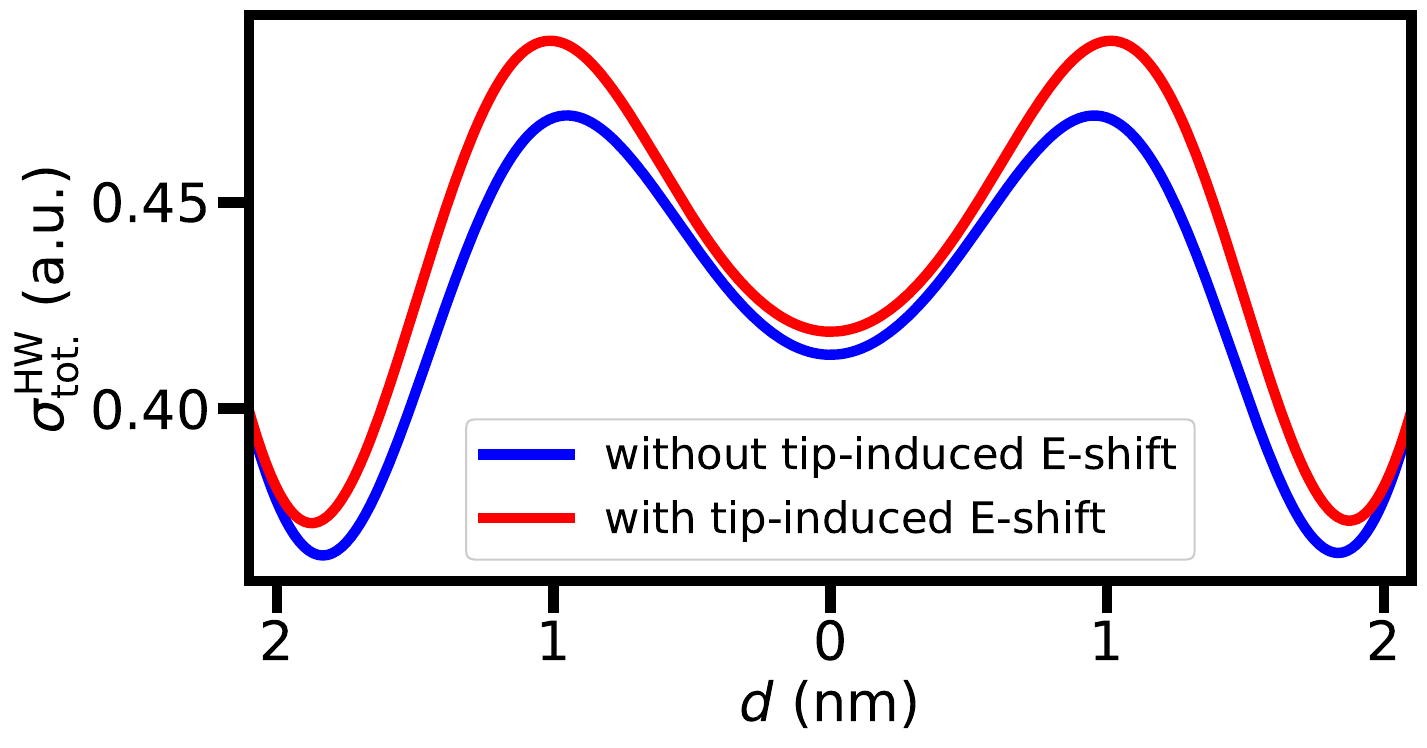}%
		\caption{\label{fig:diff_occu}
			\color{black} Simulated total surface charge density $\sigma_\mathrm{tot}^\mathrm{HW}$ for the case of unshifted (blue) and tip-shifted (red) corral states.
			The left panel shows $\sigma_\mathrm{tot}^\mathrm{HW}$ along the diagonal of the corral, vanishing at the confining walls (the boundary). 
			The right panel is a zoom into the central region, where it becomes clearer that the red curve (tip-influenced) is spatially broader than the blue one. 
			$d$ gives the distance from the center of the corral.
			This behavior mirrors the difference observed in the AFM line profiles shown in Fig.2(f) of the main text. 
			Note that only the case of downward energy shift (metal-tip) is shown here to illustrate the general principle; the opposite effect is expected for CO-tips, which raise the corral state energies and reduce their occupation.} 
	\end{figure}

}

\newpage

\section{Definition of the frequency shift contrasts and the error bars}

To determine the frequency shift contrast $C^\mathrm{\Delta f}$ at a specific height $z$, we first identified the frequency shift values at the outer extrema ($\Delta f^{\times}(x_i, y_i,z)$) and the central extremum ($\Delta f^{\Box}(x_0, y_0, z)$) within the image. In FIGs. 2(b) and (d) in the main text outer extrema are marked by blue crosses, and the central extreme point is marked by the green square. The average frequency shift of the outer extrema was calculated as $\overline{\Delta f^{\times}}(z)$. The contrast at a specific height is then given by:

\begin{equation}
C^\mathrm{\Delta f}(z) = \bigl| \overline{\Delta f^{\times}}(z) - \Delta f^{\Box}(x_0, y_0, z) \bigr|.
\end{equation}

The error in contrast is determined by the uncertainty in the frequency shift values. To calculate the error range, we considered all combinations of $\Delta f^{\times}(x_i, y_i, z)$ and $\Delta f^{\Box}(x_0, y_0, z)$.  
The maximum and minimum contrast values define the error bars shown in the plots.

\newpage

\section{Analyzing the contrast evolution of the second less dense corral}

\begin{figure}[H]
	\centering
	\includegraphics[width=0.49\textwidth]{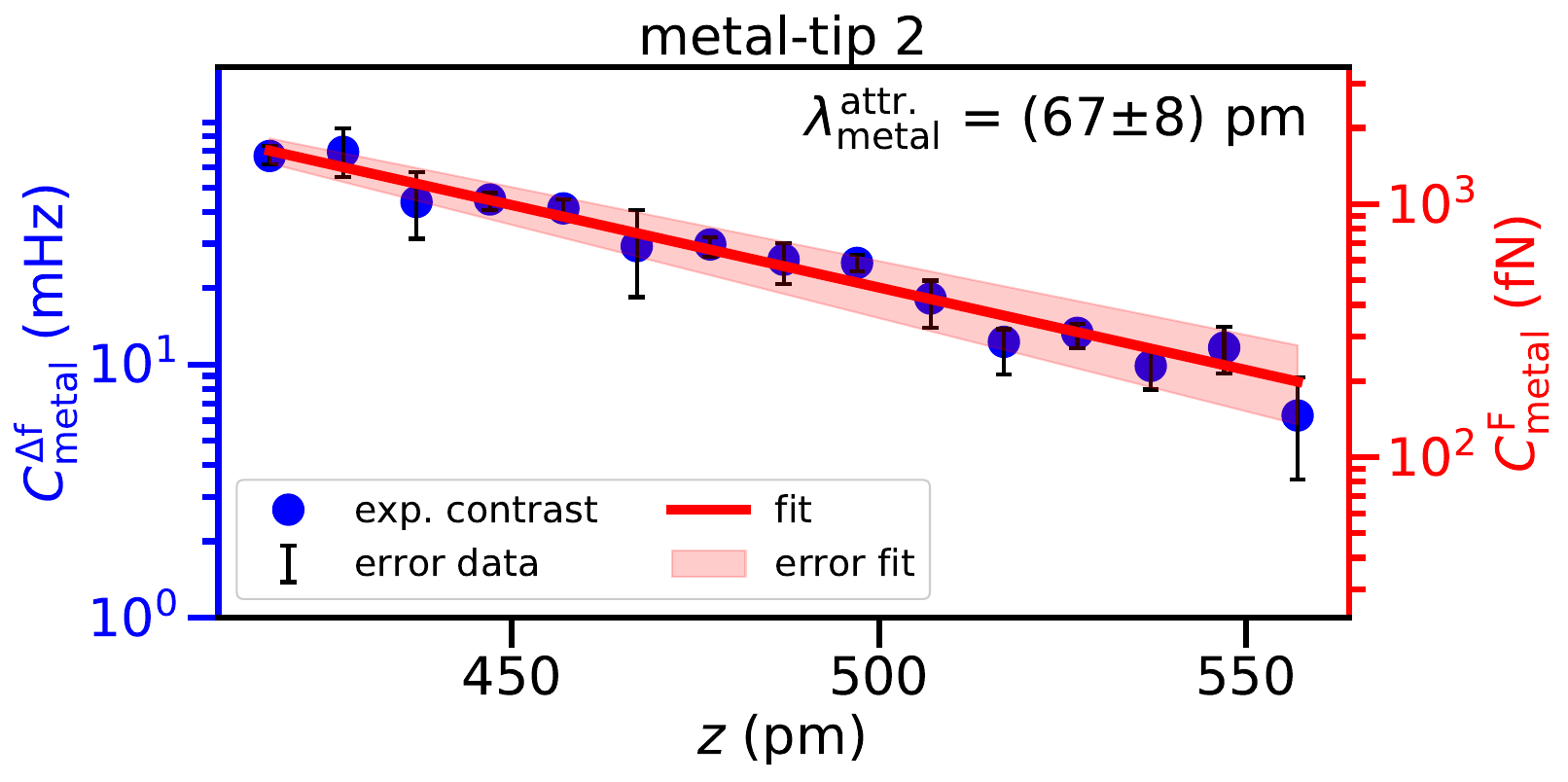}%
	\hfill
	\includegraphics[width=0.49\textwidth]{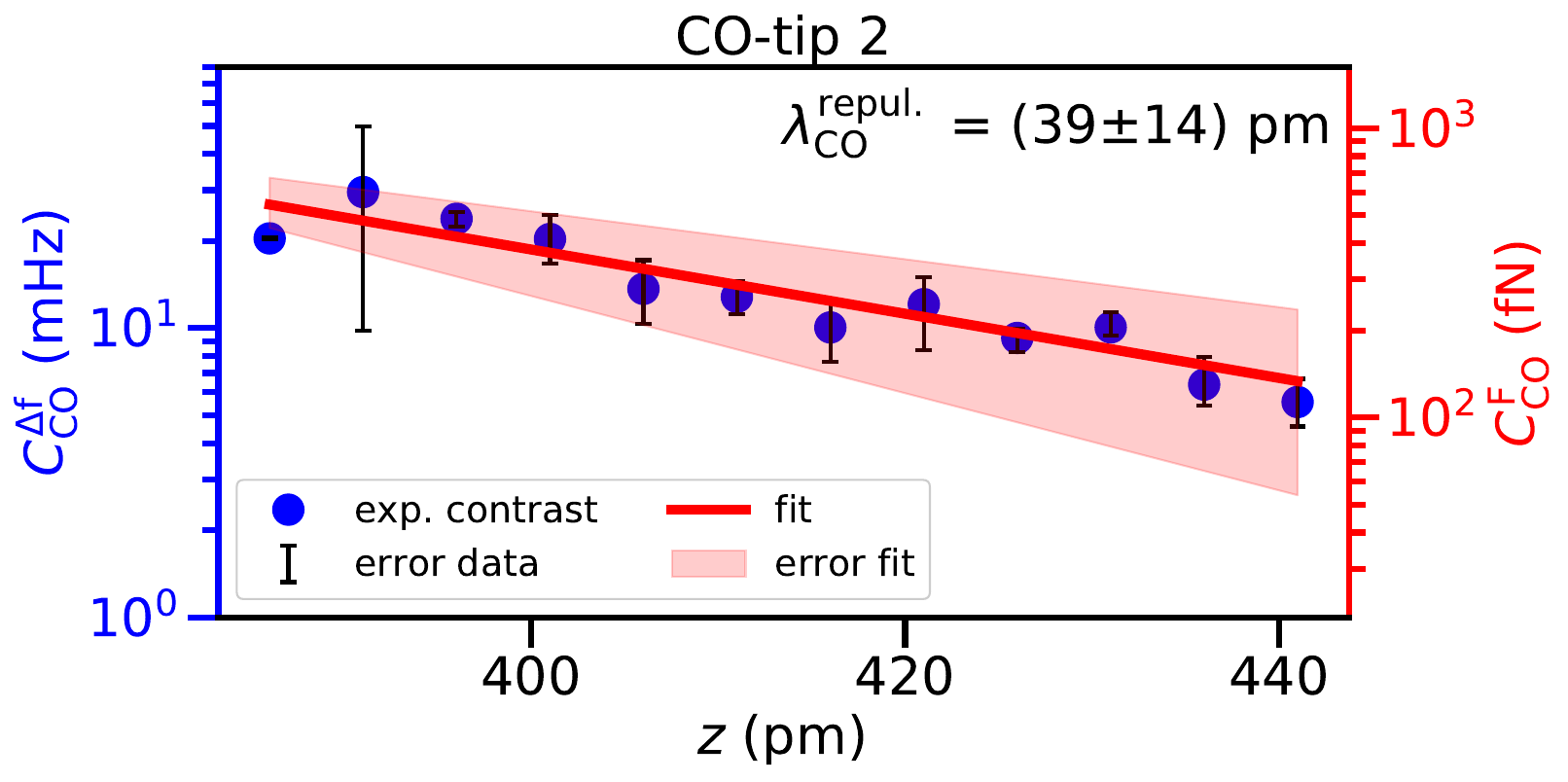}%
	\includegraphics[width=0.49\textwidth]{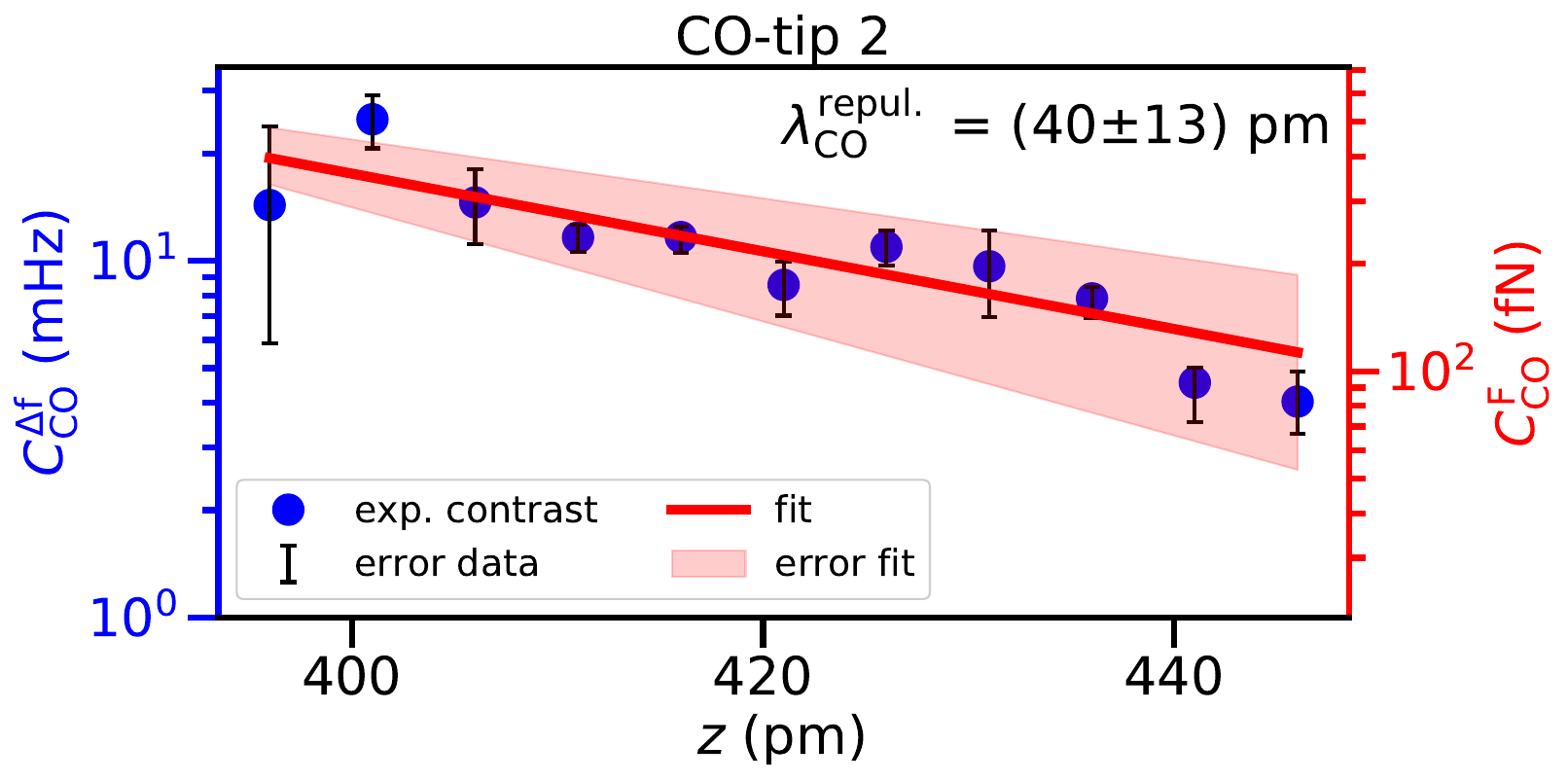}%
	\caption{\label{2nd_less_dense_corral} Contrast in $\Delta f$ and $F$, measured with CO- and metal-tip 2, as a function of vertical distance $z$ showing an exponential behavior with decay lengths of $\lambda_\mathrm{CO}^\mathrm{repul.} = (39 \pm 14)$\;pm, $\lambda_\mathrm{CO}^\mathrm{repul.} = (40 \pm 13)$\;pm , and $\lambda_\mathrm{metal}^\mathrm{attr.} = (67 \pm 8)$\;pm. To show reproducibility within one measurement set, the contrast evolution was determined twice with the same CO-tip} 
\end{figure}

Like written in the main text, to show the reproducibility of the measured decay lengths, a second equivalent less dense corral was constructed. Between the measurement set of the first less dense corral and the second, multiple sputter and annealing cycles were performed. Furthermore, the mesoscopic tip shape was altered by several hundred controlled indentations of the tip in the Cu(111) sample. In Figure \ref{2nd_less_dense_corral} this is marked with metal-tip 2 and CO-tip 2. To show the reproducibility within one measurement set, the contrast evolution was determined twice with the same CO-tip.

\newpage

\section{Modeling Pauli repulsion and chemical bond}

For modeling the interactions of the corral states with the probe tip, via an overlap integral, normalized s-wave functions for tip and sample were used: 

\begin{equation}\label{eq:swave_function}
\Psi_\mathrm{A}(\Vec{r}) = \dfrac{1}{\sqrt{\pi (\lambda_\mathrm{A}^\Psi)^3}} \mathrm{exp} \biggl(- \dfrac{|\Vec{r}-\Vec{R}_\mathrm{A}|}{\lambda_\mathrm{A}^\Psi} \biggr) \qquad \mathrm{and} \qquad \Psi_\mathrm{B}(\Vec{r}) = \dfrac{1}{\sqrt{\pi (\lambda_\mathrm{B}^\Psi)^3}} \mathrm{exp} \biggl(- \dfrac{|\Vec{r}-\Vec{R}_\mathrm{B}|}{\lambda_\mathrm{B}^\Psi} \biggr)
\end{equation}

Here, $\Vec{R}_\mathrm{A}$ and $\Vec{R}_\mathrm{B}$ are given by the centers of the respective wave functions and $\lambda_\mathrm{A}^\Psi$ and $\lambda_\mathrm{B}^\Psi$ describe the decay lengths of the wave functions. The overlap integral is then given by:

\begin{equation}
S =  \int \Psi_\mathrm{A}(\vec{r}) \Psi_\mathrm{B}(\vec{r}) \mathrm{d} \vec{r}
\end{equation}

The advantage of using s-like wave functions is, that the overlap integral is analytically solvable with $R = |\Vec{R}_\mathrm{A} - \Vec{R}_\mathrm{B}|$ being the distance between the two centers of the wave functions:

\begin{alignat}{2}\label{overlap_solution}
&S =\begin{cases} \biggl(1+\dfrac{R}{\lambda^\Psi} + \dfrac{R^2}{3 (\lambda^\Psi)^2}\biggr) \mathrm{exp}\biggl(-\dfrac{R}{\lambda^\Psi}\biggr) &\mbox{if } \lambda_\mathrm{A}^\Psi = \lambda_\mathrm{B}^\Psi = \lambda^\Psi \\ \chi [S_{1}S_{2} - S_3 S_{4}] & \mbox{if } \lambda_\mathrm{A}^\Psi \neq \lambda_\mathrm{B}^\Psi\end{cases}
\end{alignat}

with

\begin{align*}
\chi &= \dfrac{R^3}{4} \biggl( \dfrac{1}{\lambda_\mathrm{A}^\Psi}\biggr)^{3/2} \biggl( \dfrac{1}{\lambda_\mathrm{B}^\Psi}\biggr)^{3/2}\\
S_{1} &= \dfrac{(K_p^2 + 2 K_p + 2) \mathrm{exp}({-K_p})}{K_p^3}\\
S_{2} &= \dfrac{\mathrm{exp}({K_m}) - \mathrm{exp}({-K_m})}{K_m} \\
S_{3} &= \dfrac{\mathrm{exp}({-K_p})}{K_p}\\
S_{4} &= \dfrac{\mathrm{exp}({-K_m}) \left[ (K_m^2 - 2 K_m + 2) \mathrm{exp}({2 K_m}) - K_m^2 - 2 K_m - 2 \right]}{K_m^3}\\
\end{align*}

and

\begin{align*}
K_\mathrm{p} = \dfrac{R}{2\lambda_\mathrm{A}^\Psi} + \dfrac{R}{2\lambda_\mathrm{B}^\Psi}\\
K_\mathrm{m} = \dfrac{R}{2\lambda_\mathrm{A}^\Psi} - \dfrac{R}{2\lambda_\mathrm{B}^\Psi}.
\end{align*}

To use Equation (\ref{overlap_solution}) for the overlap of electron densities ($\rho$), we must consider that the electron density's decay length ($\lambda^\rho$) is half that of the corresponding wavefunction ($\lambda^\Psi$): since $|\Psi|^2 = \rho$ and $|\mathrm{exp}(-|r|/\lambda^\Psi)|^2 = \mathrm{exp}(-(2|r|)/\lambda^\Psi) = \mathrm{exp}(-|r|/\lambda^\rho)$ with $\lambda^\rho = \lambda^\Psi/2$.
Furthermore, Equation (\ref{overlap_solution}) requires additional scaling factors:
\begin{itemize}
	\item $\dfrac{1}{64 \pi} \biggl( \dfrac{1}{\lambda^\rho}\biggr)^{3/2} \biggl( \dfrac{1}{\lambda^\rho}\biggr)^{3/2}$ when $\lambda_\mathrm{A}^\rho = \lambda_\mathrm{B}^\rho = \lambda^\rho$
	\item $\dfrac{1}{64 \pi} \biggl( \dfrac{1}{\lambda_\mathrm{A}^\rho}\biggr)^{3/2} \biggl( \dfrac{1}{\lambda_\mathrm{B}^\rho}\biggr)^{3/2}$ when $ \lambda_\mathrm{A}^\rho \neq \lambda_\mathrm{B}^\rho$.
\end{itemize}

\subsection{overlap integral of electron densities to model Pauli repulsion}\label{sec:Pauli}

\begin{figure}[H]
	\centering
	\includegraphics[width=0.49\textwidth]{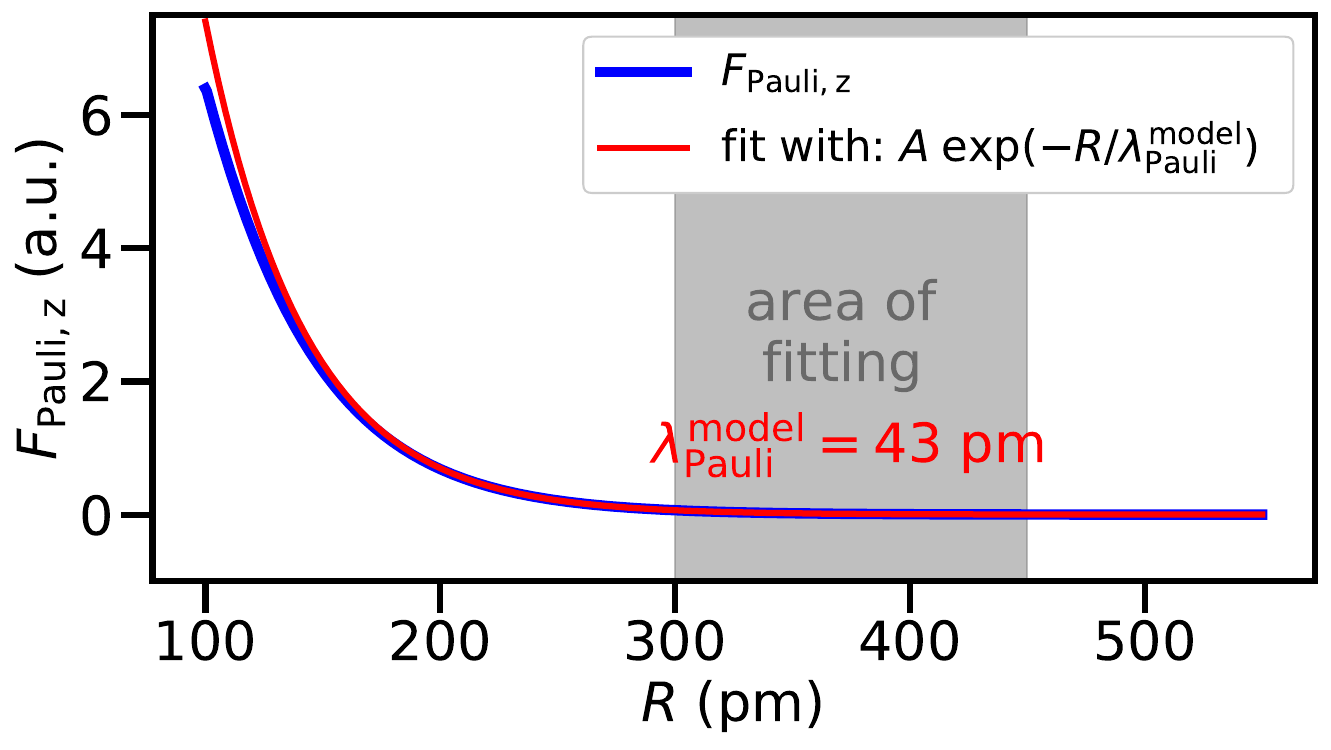}%
	\caption{\label{overlap_pauli} Repulsive force calculated by equation (\ref{Pauli}) and (\ref{overlap_solution}) when approximating the tip and sample electron density as s-like distributions (blue graph). $R$ defines the distance between tip and surface. The electron density decay lengths of tip and sample are set to $\lambda_\mathrm{sample}^\mathrm{\rho} = 42$ pm and $\lambda_\mathrm{CO-tip}^\mathrm{\rho} = 12$ pm (see main text for explanation). Fitting this curve with an exponential function (red) in the region $[300,450]$ pm (the height of the CO-tip measurements) gives a fitted decay length of $\lambda_\mathrm{Pauli}^\mathrm{model} = 43$ pm.} 
\end{figure}

The distance behavior of the potential energy of Pauli repulsion is often modeled by the overlap of electron densities. Taking the derivative in $z$-direction then gives the normal force, $F_\mathrm{Pauli,z}$ (which AFM is sensitive to). Using the s-wave approximation mentioned earlier, the Pauli repulsion force is given by:

\begin{equation}\label{Pauli}
F_\mathrm{Pauli,z} \propto - \frac{\partial}{\partial z} S_\rho = -\frac{\partial}{\partial z} \int \rho_\mathrm{tip}(\Vec{r}) \rho_\mathrm{sample}(\Vec{r}) \; \mathrm{d}\Vec{r}
\end{equation}

with the analytical solution of the overlap integral, $S_\rho$, being described by equation (\ref{overlap_solution}). Using the electron density decay lengths described in the main text ($\lambda_\mathrm{sample}^\mathrm{\rho} = 42$ pm  and $\lambda_\mathrm{CO-tip}^\mathrm{\rho} = 12$ pm) results in Figure \ref{overlap_pauli}. Fitting the modeled curve with an exponential function in the height where the CO-tip measurements were
conducted ([300, 450] pm) gives a decay length of $\lambda_\mathrm{Pauli}^\mathrm{model} = 43$ pm.

\subsection{overlap integral of wave functions to model chemical attraction}\label{sec:wave functionoverlap}
The distance behavior of the potential energy of a chemical bond is often approximated by the wave function overlap, $S_\mathrm{\Psi}$. Taking the derivative in $z$-direction again gives the normal force, $F_\mathrm{chem, z}$. To account for an attractive force a phenomenological minus sign was introduced:

\begin{equation}\label{overlap_wavefunct}
F_\mathrm{chem, z} \propto - \frac{\partial}{\partial z} (- S_\Psi) = \frac{\partial}{\partial z} \int \Psi_\mathrm{tip}(\Vec{r}) \Psi_\mathrm{sample}(\Vec{r}) \; \mathrm{d}\Vec{r}.
\end{equation}

\begin{figure}[H]
	\centering
	\includegraphics[width=0.49\textwidth]{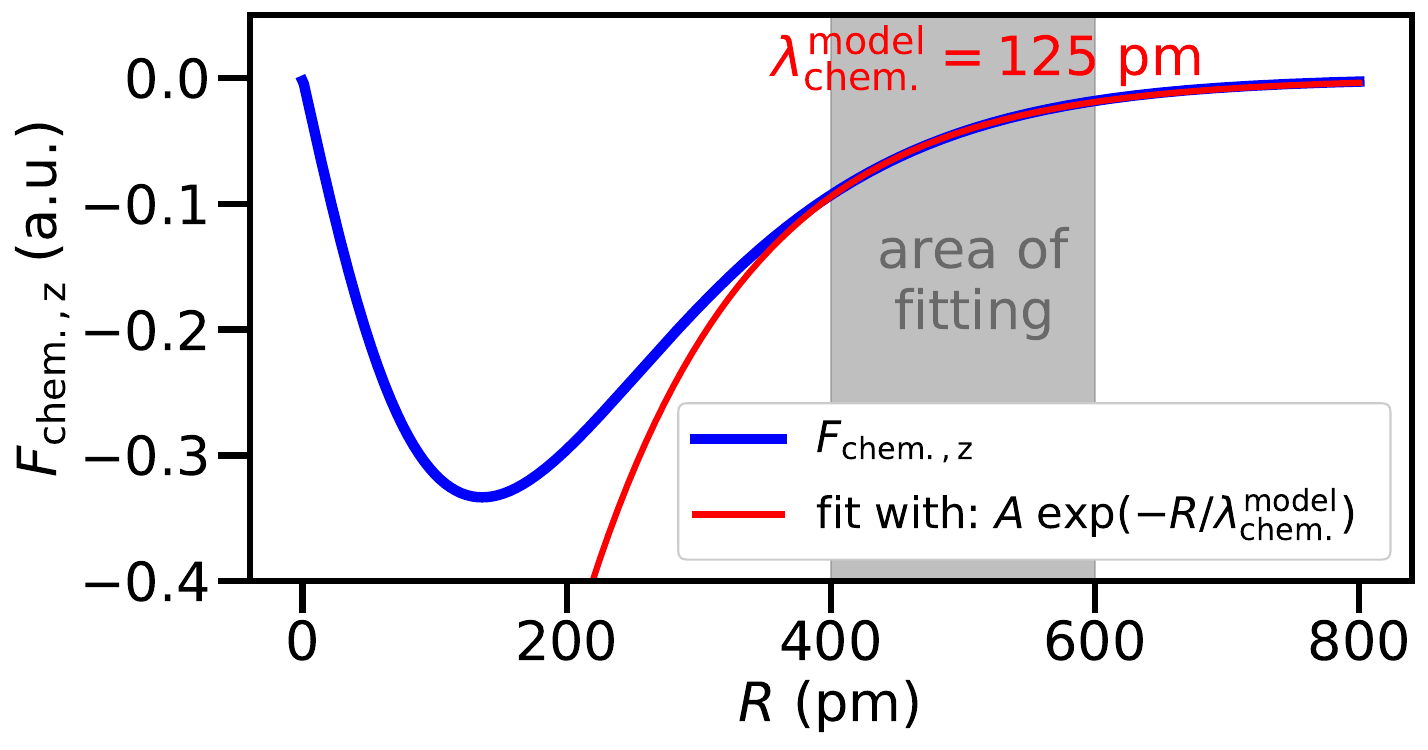}%
	\caption{\label{overlap_swaves_plot} Attractive force calculated by equation (\ref{overlap_wavefunct}) and (\ref{overlap_solution}) when approximating the tip- and sample-wave function as s-waves (blue graph). $R$ defines the distance between tip and surface. The decay lengths of tip and sample wave functions are set to $84$ pm. Fitting this curve with an exponential function (red) in the region $[400,600]$ pm (the height of the metal-tip measurements) gives a fitted decay length of $\lambda_\mathrm{chem.}^\mathrm{model} = 125$ pm.} 
\end{figure}

Tip and corral wave functions are modeled by s-waves (see equation (\ref{eq:swave_function})). The decay lengths of tip and sample wave functions both depend on the work function (see main text for explanation) and are: $\lambda_\mathrm{sample}^\mathrm{\Psi} = \lambda_\mathrm{tip}^\mathrm{\Psi} = 84$ pm. 
Fitting Eq. (\ref{overlap_wavefunct}) with an exponential function in the height where the metal-tip measurements were conducted ($[400,600]$ pm) gives a decay length of $125$ pm. The plot is shown in FIG. \ref{overlap_swaves_plot}.

\subsection{overlap integral of electron densities to estimate the Pauli repulsion decay length for a metal-tip}\label{sec:densityoverlap}
{\color{black}
	Like mentioned in the main text the repulsive force regime is not accessible with a metal-tip. Assuming the description of Pauli repulsion with an electron density overlap integral also holds true for the metal-tip, one can estimate the repulsive decay length. Using the s-wave approximation the Pauli repulsion force is given by:
}

\begin{equation}\label{overlap_density_metal}
F_\mathrm{Pauli, z}^\mathrm{metal} \propto -\frac{\partial}{\partial z} (-S_\rho) =  \int \rho_\mathrm{tip}(\Vec{r}) \rho_\mathrm{sample}(\Vec{r}) \; \mathrm{d}\Vec{r}.
\end{equation}

\begin{figure}[H]
	\centering
	\includegraphics[width=0.49\textwidth]{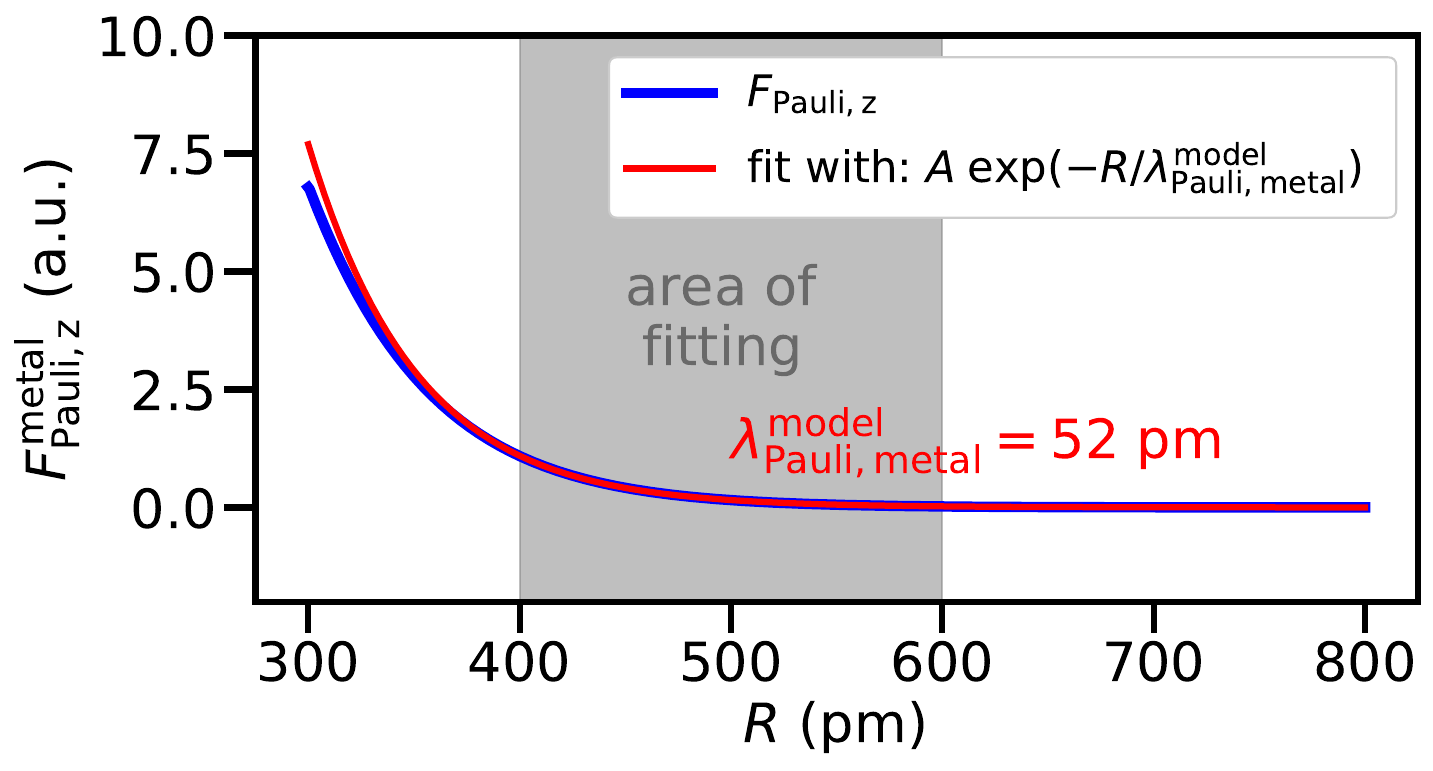}%
	\caption{\label{model_repul_metal} Repulsive force calculated by equation (\ref{Pauli}) and (\ref{overlap_density_metal}) when approximating the tip and sample electron density as s-like distributions (blue graph). $R$ defines the distance between tip and surface. The electron density decay lengths of tip and sample are set to $\lambda_\mathrm{tip}^\mathrm{\rho} = \lambda_\mathrm{sample}^\mathrm{\rho} = 42 pm$ (see main text for explanation). Fitting this curve with an exponential function (red) gives a rough estimate of the decay length of $\lambda_\mathrm{Pauli, metal}^\mathrm{model} = 52$ pm.} 
\end{figure}

Since $|\Psi|^2 = \rho$, the decay lengths of tip and sample electron densities ($\lambda_\mathrm{tip}^\mathrm{\rho}$ and $\lambda_\mathrm{sample}^\mathrm{\rho}$)  both depend on the work function and are half of the wave function decay lengths: $\lambda_\mathrm{tip}^\mathrm{\rho} = \lambda_\mathrm{sample}^\mathrm{\rho} = \lambda_\mathrm{tip}^\mathrm{\Psi}/2 = \lambda_\mathrm{sample}^\mathrm{\Psi}/2 = 84/2\; \mathrm{pm} = 42\;\mathrm{pm}$. The overlap integral of electron densities (Eq. (\ref{overlap_density_metal})) is plotted in Figure \ref{model_repul_metal}. Fitting with an exponential function in the height where the metal-tip measurements were conducted ($[400,600]$ pm) gives a decay length of $\lambda_\mathrm{Pauli, metal}^\mathrm{model} = 52$ pm.

\subsection{Influence of the tip decay length on the modeled decay length}\label{sec:influence_on_decay}

To provide an intuitive understanding of how the decay length of the force depends on the decay lengths of tip and sample, we performed a systematic analysis based on the electron density overlap model introduced earlier. Using the analytical form of the overlap integral (Eq. (\ref{overlap_solution})), we fixed the decay length of the sample electron density to $\lambda_\mathrm{sample}^\rho = 42$ pm and varied the tip decay length $ \lambda_\mathrm{tip}^\rho $ from $1$ pm to $100$ pm. For each combination, the resulting force curve was calculated and fitted with an exponential function in the height-range $[400, 600]$ pm, similar to the main analysis.

\begin{figure}[H]
	\centering
	\includegraphics[width=0.49\textwidth]{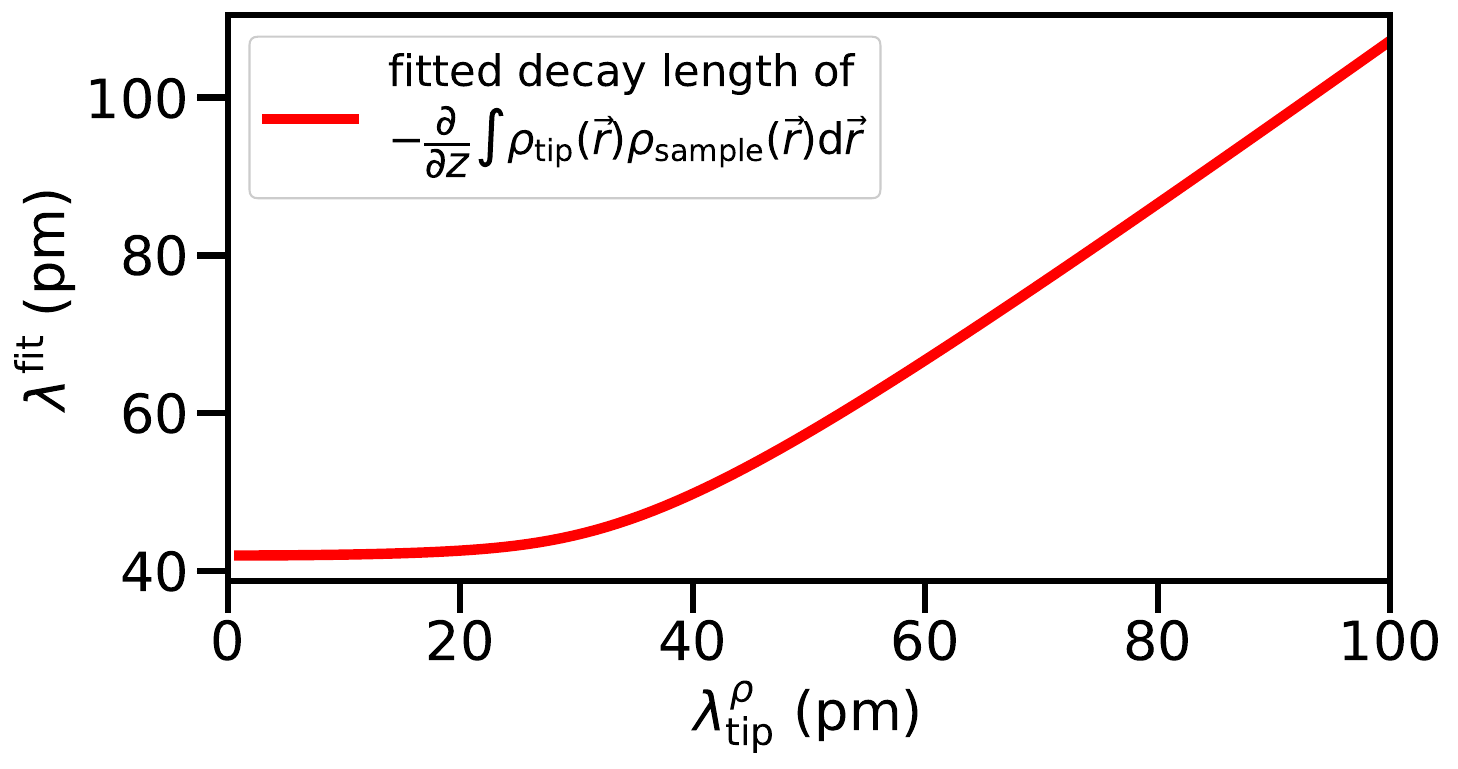}%
	\caption{\label{decay_lengths} Simulated dependence of the fitted decay length $\lambda_\mathrm{fit}$ on the tip decay length $\lambda_\mathrm{tip}^\rho$. The electron density of the sample was kept fixed at $\lambda_\mathrm{sample}^\rho = 42$ pm, while $\lambda_\mathrm{tip}^\rho$ was varied between $1$ pm and $100$ pm. For each combination, the resulting force profile was fitted with an exponential function in the range $[400, 600]$ pm to extract $\lambda_\mathrm{fit}$. The transition from a sample-dominated (for $\lambda_\mathrm{tip}^\rho \lesssim 30$ pm) to a tip-influenced regime (for $\lambda_\mathrm{tip}^\rho \gtrsim 30$ pm) illustrates the convolution-like nature of the force decay.} 
\end{figure}

The extracted decay lengths $\lambda_\mathrm{fit}$ are plotted as a function of $\lambda_\mathrm{tip}^\rho$ in Fig. \ref{decay_lengths}. The result shows that for short tip decay lengths, $\lambda_\mathrm{fit}$ is dominated by the sample’s decay length and remains nearly constant. As $\lambda_\mathrm{tip}^\rho$ increases, it begins to influence the effective decay behavior, and for sufficiently large values, $\lambda_\mathrm{fit}$ increases linearly with $\lambda_\mathrm{tip}^\rho$. This smooth transition from sample-dominated to tip-influenced decay illustrates the convolution-like nature of the interaction and explains, for example, why the modeled decay length for Pauli repulsion in the CO-tip chase (section \ref{sec:Pauli}) is only slightly larger than the sample decay length: The very short decay of the CO-tip electron density ($12$ pm) contributes minimally to the spatial range of the interaction.

\subsection{Justification of the s-wave approximation}

To simplify the overlap integral calculation, we approximated both the tip and sample with s-like wave functions and electron density distributions. While this approximation is valid for the tip, its applicability to the sample, given the sample's flat geometry, may seem less intuitive. However, the s-wave approximation for the sample is justified by its computational efficiency and the close agreement with more complex calculations. To assess the approximation's accuracy, we numerically calculated the overlap integral (similar to section \ref{sec:densityoverlap}) between an s-like tip electron distribution and an only z-dependent sample electron distribution:

\begin{equation}\label{eq:real_shockley_density}
\rho_\mathrm{sample}(z) \propto \mathrm{exp}\biggl(- \dfrac{|z|}{\lambda_\mathrm{sample}^\mathrm{\rho}}\biggr).
\end{equation}

\begin{figure}[H]
	\centering
	\includegraphics[width=0.49\textwidth]{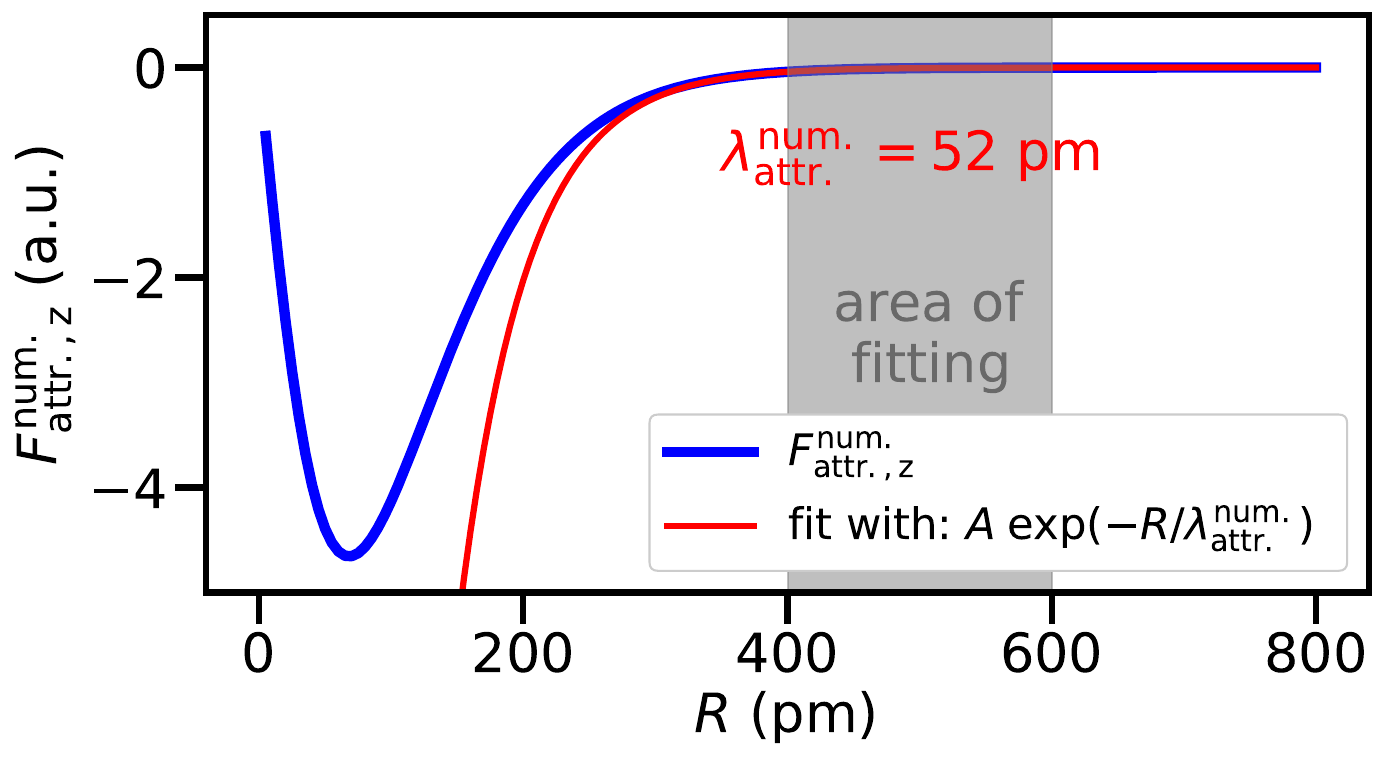}%
	\caption{\label{real_shockley_decay} Comparison of the numerically calculated overlap integral (blue curve) with an exponential fit (red line) to justify the s-wave approximation. $R$ defines the distance between tip and surface. The numerical calculation uses an s-like tip electron distribution and a z-dependent electron distribution for the sample. The fit, performed between 400 pm and 600 pm, gives a decay length of 52 pm.} 
\end{figure}

In this calculation, we used $\lambda_\mathrm{tip}^\mathrm{\rho} = \lambda_\mathrm{sample}^\mathrm{\rho} = 42$ pm. The electron density of the sample was modeled as homogeneous in the x-y plane within a 1 nm x 1 nm area (discretized with a 5 pm pixel size) and varied exponentially in the z-direction (see eq. (\ref{eq:real_shockley_density})) from -200 pm to 900 pm, with $z = 0$ defined as the surface plane. The tip was positioned at various distances from the surface, ranging from 0 pm to 800 pm, and the 3D overlap integral was calculated for each position. The results of this numerical integration are shown as the blue curve in Figure \ref{real_shockley_decay}.

To compare with the analytical s-wave result, we fitted this numerical curve with an exponential function over the range of 400 pm to 600 pm (consistent with the analysis in section \ref{sec:densityoverlap}). The fitted decay length is $\lambda_\mathrm{attr.}^\mathrm{num.}= 52$ pm (red line in Figure \ref{real_shockley_decay}). This is the same as the decay length of 52 pm obtained using the s-wave approximation in Section \ref{sec:densityoverlap}.

\newpage

{\color{black}
	\section{Correction to the fitted attractive decay length in a Morse-type potential}
	The Morse-potential is commonly used to describe atom–atom interactions. Its corresponding force function consists of two exponential terms:
	\begin{equation}
	F_\mathrm{Morse}(x) = A \exp \left(- \frac{x}{\lambda_1} \right) - B \exp \left(- \frac{x}{\lambda_2} \right) = \exp \left(- \frac{x}{\lambda_1} \right) \left( A - B \exp \left[- x \frac{\lambda_1 - \lambda_2}{\lambda_1 \lambda_2} \right] \right),
	\end{equation}
	where $A$ and $B$ are positive prefactors, $\lambda_1$ is the decay length of the repulsive component, and $\lambda_2$ is the decay length of the attractive component. The measurements in this work were performed at a position $x = x_0 + \epsilon$, where $x_0$ is the base measurement height and $\epsilon$ is the relative displacement. Inserting this into the force expression yields:
	\begin{equation}
	F_\mathrm{Morse}(x_0 + \epsilon) = \exp \left( - \frac{x_0 + \epsilon}{\lambda_1} \right) \left( A - B \exp \left[- (x_0 + \epsilon) \frac{\lambda_1 - \lambda_2}{\lambda_1 \lambda_2} \right] \right).
	\end{equation}
	As described in the main text, the full force-curve is assumed to follow a Morse-type interaction. However, in the experiment we fit the data with a single exponential. To make both descriptions comparable, we express this situation as:
	\begin{equation} \label{eq:fit_vs_morse}
	F_\mathrm{Morse}(x_0 + \epsilon) = \exp \left( - \frac{x_0 + \epsilon}{\lambda_1} \right) \left( A - B \exp \left[ - (x_0 + \epsilon) \frac{\lambda_1 - \lambda_2}{\lambda_1 \lambda_2} \right] \right)
	\approx -C \exp \left( - \frac{x_0 + \epsilon}{\lambda_\mathrm{fit}} \right).
	\end{equation}
	Here, the left-hand side describes the actual Morse-type force, and the right-hand side is the exponential function used for fitting. The constant $C>0$ is the amplitude of the fitted function. The goal is to determine how the fitted decay length $\lambda_\mathrm{fit}$ relates to the true attractive decay length $\lambda_2$.
	Since the measured forces with a metal-tip are clearly attractive ($F<0$), we multiply both sides by $-1$ and apply the natural logarithm:
	\begin{equation}\label{eq:compare_to_fit}
	\ln \left( -F_\mathrm{Morse}(x_0 + \epsilon) \right) = - \frac{x_0 + \epsilon}{\lambda_1} + \ln \left( -A + B \exp \left[ - (x_0 + \epsilon) \frac{\lambda_1 - \lambda_2}{\lambda_1 \lambda_2} \right] \right) \
	\approx \ln(C) - \frac{x_0 + \epsilon}{\lambda_\mathrm{fit}}.
	\end{equation}
	To simplify the Morse-curve, a first order Taylor expansion of the ln-term around $\epsilon=0$ is performed:
	\begin{align}\label{eq:taylor}
	\begin{split}
	\ln \left( -F_\mathrm{Morse}(x_0 + \epsilon) \right) &= - \frac{x_0}{\lambda_1} - \frac{\epsilon}{\lambda_1} + \ln \left( -A + B \exp \left[ -x_0 \frac{\lambda_1 - \lambda_2}{\lambda_1 \lambda_2} \right] \right) \\
	&\quad + \frac{B}{A \exp \left( x_0 \frac{\lambda_1 - \lambda_2}{\lambda_1 \lambda_2} \right) - B} \frac{\lambda_1 - \lambda_2}{\lambda_1 \lambda_2} \epsilon + \mathcal{O}(\epsilon^2).
	\end{split}
	\end{align}
	This can now be compared to the logarithmic form of the fitted function (right hand side of Eq. (\ref{eq:compare_to_fit})). We additionally introduce a general ratio $\mu = \lambda_2/\lambda_1$ with $\mu >1$, which allows us to rewrite expression (\ref{eq:taylor}) as:
	\begin{align}
	\ln \left( -F_\mathrm{Morse}(x_0 + \epsilon) \right) = \left( \frac{\mu}{\lambda_2} - \kappa(x_0) \right) \epsilon + \text{offset} + \mathcal{O}(\epsilon^2),
	\end{align}
	with the correction term
	\begin{align}
	\kappa(x_0) = \frac{B}{A \exp \left( x_0 \frac{1 - \mu}{\lambda_2} \right) - B} \frac{1 - \mu}{\lambda_2}.
	\end{align}
	Please note, that all terms independent on $\epsilon$ were combined to an offset. By comparing coefficients of $\epsilon$, we obtain the effective fitted decay length:
	\begin{equation}
	\frac{1}{\lambda_\mathrm{fit}} = \frac{\mu}{\lambda_2} - \kappa(x_0).
	\end{equation}
	This shows that $\lambda_\mathrm{fit}$ is not equal to $\lambda_2$, but is influenced by the correction term $\kappa(x_0)$, which depends on the measurement height $x_0$.
	
	In the limit of a large tip–sample separation ($x_0 \rightarrow \infty $), the repulsive exponential becomes negligible. The correction term becomes:
	\begin{align}
	\kappa(x_0 \rightarrow \infty) = - \frac{1 - \mu}{\lambda_2},
	\end{align}
	which leads to:
	\begin{align}
	\frac{1}{\lambda_\mathrm{fit}} = \frac{\mu}{\lambda_2} + \frac{1 - \mu}{\lambda_2} = \frac{1}{\lambda_2} \quad \Rightarrow \quad \lambda_\mathrm{fit} = \lambda_2.
	\end{align}
	This confirms that the fitted decay length matches the true attractive decay length only in the limit of large separation.
	The other limiting case is the minimum of the Morse potential at:
	\begin{equation}
	x_\mathrm{min} = - \frac{\lambda_2}{1 - \mu} \ln \left( \frac{\mu A}{B} \right).
	\end{equation}
	At this point, the correction term becomes:
	\begin{equation}
	\kappa(x_\mathrm{min}) = \dfrac{\mu}{\lambda_2},
	\end{equation}
	which leads to:
	\begin{equation}
	\dfrac{1}{\lambda_\mathrm{fit}} = 0 \quad \Rightarrow \quad \lambda_\mathrm{fit} = \infty.
	\end{equation}
	$\kappa$ is monotonic in this range and therefore this defines the boundaries for the fitted decay length:
	\begin{equation}
	\lambda_\mathrm{fit} \in [\lambda_2, \infty),
	\end{equation}
	and yields the central result:
	\begin{equation}
	\lambda_\mathrm{fit} \geq \lambda_2.
	\end{equation}
	
}


\begin{thebibliography}{92}
\bibitem [{foo({\natexlab{a}})}]{footnote3}%
\BibitemOpen
\bibinfo {note} {Although both measurement sets were performed on the same
	sample, multiple sputtering and annealing cycles were carried out between the
	two sessions. Furthermore, the mesoscopic configuration of the tip was
	different. During the interval between the two measurement runs, the tip apex
	was modified several hundred times through slight and controlled indentations
	into the copper sample.}\BibitemShut {Stop}%
\bibitem [{foo({\natexlab{b}})}]{footnote4}%
\BibitemOpen
\bibinfo {note} {Equivalent to the calculation of the decay length of the
	quantum corral states the tips decay length is approximated by
	$\lambda_\mathrm{tip} = \sqrt{\hbar^2/(2 m_\mathrm{e}
		(\Phi+440\,\mathrm{meV}) } = 84$ pm, where $m_\mathrm{e}$ is the electron
	mass, and $\Phi = 4.95$ eV the work function of copper. In this approximation
	the decay lenghth of the tip is calculated at the onset of the Shockley
	surface state, which is $440$ meV below the Fermi level. Between the the band
	edge of the Shockley surface state and the Fermi energy
	$\lambda_\mathrm{tip}$ only varies by 5\%. This variation is neglected here
	and will not change the interpretation.}\BibitemShut {Stop}%
\end{thebibliography}
\end{document}